\documentclass[12pt]{article}

\usepackage[T1]{fontenc}
\usepackage[utf8]{inputenx}

\usepackage[hyphens]{url}
\usepackage[margin=1in]{geometry}
\usepackage{amsmath,amssymb}
\usepackage{booktabs}
\usepackage{graphicx}
\usepackage{setspace,lipsum}%
\usepackage{tikz}
\usepackage{comment}
\usepackage{textcomp}
\usepackage{caption}
\usepackage{longtable}
\usetikzlibrary{positioning,arrows,shapes,trees}
\usepackage{threeparttablex}
\usepackage{longtable}
\usepackage{epstopdf}
\usepackage{comment}
\usepackage{siunitx}
\usepackage{xcolor}
\usepackage{natbib}
\usepackage{algorithm}
\usepackage{algpseudocode}
\usepackage{caption}
\usepackage{subcaption}
\setcitestyle{authoryear}

\usepackage[english]{babel}
\DeclareMathOperator*{\argmin}{arg\,min}
\newcommand{\vect}[1]{\ensuremath{\boldsymbol{\mathbf{#1}}}}
\newcommand{\mat}[1]{\ensuremath{\boldsymbol{\mathbf{#1}}}}

\DeclareUnicodeCharacter{2212}{-}
\begin{document}

\title{Maximum Likelihood Estimation of Differentiated Products Demand Systems}
\author{Greg Lewis \\ Microsoft Research  \and  Bora Ozaltun\thanks{Corresponding author. Email: \texttt{ozaltun@mit.edu}. We are grateful to Ariel Pakes for useful conversations early on in this project, and to Surya Ierokomos for excellent research assistance.} \\ Microsoft Research  \and Georgios Zervas \\ Boston University}
\date{\today}
\maketitle
\onehalfspacing

\begin{abstract}
We discuss estimation of the differentiated products demand system of \citet{berry1995automobile} (BLP) by maximum likelihood estimation (MLE). 
We derive the maximum likelihood estimator in the case where prices are endogenously generated by firms that set prices in Bertrand-Nash equilibrium.  
In Monte Carlo simulations the MLE estimator outperforms the best-practice GMM estimator on both bias and mean squared error when the model is correctly specified. 
This remains true under some forms of misspecification. 
In our simulations, the coverage of the ML estimator is close to its nominal level, whereas the GMM estimator tends to under-cover. 
We conclude the paper by estimating BLP on the car data used in the original \citet{berry1995automobile} paper, obtaining similar estimates with considerably tighter standard errors.
\end{abstract}

\newpage

\section{Introduction}
\label{sec:intro}

Demand estimation is a fundamental tool in applied economics.  One widely used demand model for differentiated goods markets is \citet{berry1995automobile} (BLP), which builds on the earlier work of \citet{mcfadden86} concerning the mixed logit.  
The BLP model is traditionally estimated by generalized method of moments (GMM), but the estimates are often imprecise.  
It has long been known that the estimation performance is dramatically improved by adding a model of the supply side, typically one in which firms engage in Bertrand-Nash pricing.  A small but influential literature has documented other ways of improving estimation performance: replacing a step that inverts from market shares to mean utilities with a set of constraints on the difference between actual and predicted market shares (the MPEC approach of \cite{dube2012improving}), using empirical likelihood instead of GMM \citep{conlon2013empirical}), choosing the right instruments (\cite{reynaert2014improving,gandhi2019measuring}) and using high-quality software (\cite{conlon2020best}).    

This paper adds to this literature by revisiting an old idea: estimating BLP by maximum likelihood.  
The theoretical advantages of doing so are clear.
First, maximum likelihood is  statistically efficient, which translates into more precise estimates.
Second, maximum likelihood does not require the researcher to choose their instruments. 
In fact, provided the model is correctly specified, one can think of MLE as a magical procedure that automatically performs GMM with the optimal instruments.
This may be important in practice, for two reasons.
First, choosing good instruments can be hard, as attested to by the papers on this topic cited above.
Second, BLP is often used for demand analysis in merger cases, and in such cases any degrees of freedom, such as instrument choice, can be exploited by competing expert witnesses in their analysis of the case.

There are, however, some drawbacks. Both price and quantity are endogenous, and so a well-specified likelihood requires modeling how both price and quantity arise, necessitating models of both demand and supply.
This requires committing to some model of pricing.
Moreover, the researcher needs to specify a distribution for the demand and cost shocks. \
We investigate these trade offs here.  

The paper proceeds in three parts.
First, we outline the BLP model and derive the maximum likelihood estimator (MLE) under Bertrand-Nash pricing.
Somewhat surprisingly, this appears to be new, as the prior literature we review below has assumed that the demand error enters the pricing equation linearly, which is inconsistent with Bertrand-Nash.  

The second part of the paper investigates the performance of the MLE relative to the best practice in implementing GMM (for best practice, we follow \citet{conlon2020best}).  
We start with a variety of well-specified scenarios, in which we would expect MLE to outperform GMM.
We find that it does, and quite handily - bias and mean squared error are smaller, standard errors are tighter, and coverage is more precise.
We then consider three scenarios in which the model is wrong in some way.
The first tests the robustness of MLE to a different error structure, namely Laplace errors. We find that the MLE continues to outperform the GMM benchmark.
The second considers a misspecified supply side, where estimation assumes costs are log-linear rather than linear in characteristics.
Here we find that both GMM and MLE perform poorly in estimating the demand parameters, as expected, but MLE performs worse.
Finally, we consider a case where the ownership matrix is mis-specified, which we intend as reduced form shorthand for a mis-specification of the game being played on the supply side.
The results of GMM and MLE are comparable in this case.

In the final part of the paper, we replicate BLP, using their original data, and building on the replication exercise done in \cite{andrews2017measuring} (AGS).  
We compare the MLE estimates to a series of benchmarks: the original estimates in BLP, the replication results of AGS, and the best practice estimates of \cite{conlon2020best} (CG). 
The MLE estimates are very similar to the GMM estimates both in terms of the parameters and implied own price elasticities.
Notably, the price coefficient estimate is close to the best practice of CG, unlike BLP and AGS.
In addition the standard errors are substantially smaller; often less than 25\% on a typical parameter estimate.
This performance is ``out of the box''; unlike the GMM estimators, we didn't have to choose our set of instruments.

Much of the work in this paper lies in the details. 
The likelihood requires computation of a Jacobian term, a large matrix which we analytically derive in the appendix using the implicit function theorem.
The Jacobian of the likelihood in turn includes terms from the Hessian of the share function.
Because numerical errors can easily accumulate 
we take extra care to implement state of the art computational methods. We do this by utilizing the PyTorch framework developed in \cite{paszke2019pytorch}.
One particular feature of PyTorch that helps with optimizing is Automatic Differentiation (AD). For optimizing the objective, finite difference methods can lack speed and reliability compared to analytic derivatives. AD exploits the chain rule and the fact that computer code is made up of elementary mathematical operations to compute gradients that are as precise as analytically derived gradients.

Taken together, these simulation and replication exercises suggest that MLE is a useful addition to the set of tools available to estimate BLP, allowing researchers to avoid choosing moment conditions, and having a substantial advantage in statistical precision, at least where the researcher is willing to commit to a model of the supply side.

\paragraph{Related literature.}
The idea of estimating differentiated products demand systems by maximum likelihood is by no means new.
With individual choice data, it is in fact common (see e.g. \citet{honka2017advertising} and \citet{abaluck2021consumers} for more recent examples).
The marketing literature has also considered using maximum likelihood with aggregate data.
\cite{jiang2009bayesian} suggests a Bayesian Markov Chain Monte Carlo (MCMC) approach for demand estimation in the absence of endogenous prices.
They find in simulations that their estimates have lower mean squared error.
In an extension of their main model, they add a linear supply side pricing equation and suggest how instrumental variables may be used in the presence of endogeneity.
\cite{park2009simulated} perform a similar analysis, using maximum likelihood rather than a Bayesian approach, and again insisting on a linear pricing equation and bivariate normal errors.
Finally, in an appendix to their paper, \citet{dube2012improving} once again employ a linear pricing equation when considering the performance of a maximum likelihood estimator of BLP.
Our paper differs from these prior contributions in that we set up the supply-side pricing equation to be consistent with Bertrand-Nash pricing, leading to an equation in which the demand error enters non-linearly.  


\section{Model}
\label{sec:model}

Our model closely follows~\cite{berry1995automobile}.  
Our exposition is deliberately concise, and we refer the reader to~\cite{berry1995automobile} for additional details. 
We use $i$ to index individuals, $j$ for products, and $t$ for markets. 
We present our full notation in Table \ref{model:var_def}. 

\paragraph{Utility.} An individual $i$ buying product $j$ in market $t$ obtains utility equal to:
\begin{equation}
    \label{eq:m1}
    u_{ijt} = \vect{x}_{jt}' \vect{\beta}_i - \alpha p_{jt} +\xi_{jt} + \epsilon_{ijt},
\end{equation}
where $\vect{x}_{jt}$ represents the observable characteristics of a product, $p_{jt}$ is the associated price of a product, $\xi_{jt}$ is a scalar 
unobservable product characteristic, $\delta_{jt}$ is the product's mean utility, and $\epsilon_{ijt}$ is an idiosyncratic shock that is i.i.d. and follows a type-1 extreme value distribution. 
Product characteristics are allowed to have heterogeneous effects on individuals' utilities through $\vect{\beta}_i$. 
Each observable characteristic coefficient $k$ follows $\beta_{ik} = \beta_k + \tilde{\beta}_{ik}$, where $\tilde{\beta}_{ik}$ is an i.i.d. random variable drawn from $N(0, \sigma_{\beta})$. 
Mean (across consumers) utility takes the form:
\begin{equation}
  \label{eq:demand_reg}
    \delta_{jt} = \vect{x}_{jt}' \vect{\beta} - \alpha p_{jt} + \xi_{jt},
\end{equation}
and thus individual utility can be re-written as:
\begin{equation}
    \label{eq:m2}
    u_{ijt} = \delta_{jt} + \vect{x}_{jt}' \vect{\tilde{\beta}}_{i} + \epsilon_{ijt}.
  \end{equation}
 
\paragraph{Demand.} Individual consumers make product choices that maximize their utility. 
The observed market-level shares of each product are the result of aggregating these individual-level decisions.
The assumption on the idiosyncratic shock $\epsilon_{ijt}$ and the existence of an outside option where $u_{0t}=0$ yields market shares:
\begin{equation}
  \label{eq:share_int}
  s_{jt} = \int \frac{e^{\delta_{jt} + \vect{x}_{jt}' \vect{\tilde{\beta}}_{i}}}{1 + \sum_{j' \in \mathcal{J}_t} e^{\delta_{j't} + \vect{x}_{j't}' \vect{\tilde{\beta}}_{i}}} dF(\tilde{\beta_i}|\sigma_{\beta}).
\end{equation}

\paragraph{Supply.} For the supply side of the market, we assume a marginal cost structure of:
\begin{equation}
  \label{eq:supply_reg}
  c_{jt} = \vect{w}_{jt}'\vect{\gamma} + u_{jt},
\end{equation}
where $c_{jt}$ is the marginal cost of producing a product, $\vect{w}_{jt}$ is a vector of characteristics and $u_{jt}$ is a latent scalar supply side cost shock.
Firm profits are given by:
\begin{equation}
  \label{eq:supply}
    \pi_{ft} = \max_{ p_{ft}} \sum_{j \; \in \; \mathcal{J}_{ft}}(p_{jt} - c_{jt}) s_{jt}.
\end{equation}
As in BLP, we assume that firms have the objective of maximizing their profits in a given market across all of their products.
This leads to the following first order conditions (FOCs):
\begin{equation}
  \label{eq:supply_foc}
    	\frac{\partial \pi_{ft}}{\partial p_{jt}} =  s_{jt} + \sum_{j' \; \in \; \mathcal{J}_{ft}}(p_{j't} - c_{j't})\frac{\partial s_{j't}}{\partial p_{jt}},
\end{equation}
where $\pi_{ft}$ is the profit of firm $f$ in market $t$.
This can be written in matrix form as:
\begin{equation}
      \vect{c}_{t} = \vect{p}_{j} - (\mat{O}_t  \mat{J}_{t}^{s,p})^{-1} \vect{s}_t,
\end{equation}
where $J^{s,p}_{t}[j, j'] = \frac{\partial s_{j'}}{\partial p_j}$ and $O_t$ is the ownership matrix: 
\begin{equation}
O_t = \begin{cases} 
  1 \; \text{if }j \wedge j' \; \in \mathcal{J}_{ft},\\
  0 \; \text{otherwise.}
  \end{cases}
\end{equation}

Given $\{\vect{p}_{t}, \vect{s}_{jt}\}$ and a set of parameters, Eq.~\ref{eq:supply_foc} can be used to estimate marginal costs. 
These marginal costs can be used to estimate the parameters of Eq.~\ref{eq:supply_reg}:
\begin{equation}
  \label{eq:supply_reg_2}
  \vect{c}_t = \vect{p}_t - (\vect{O}_t \vect{J}_{t}^{s,p})^{-1}  \vect{s}_t = \vect{w}_{t} \vect{\gamma} + \vect{u}_{t}.
\end{equation}

\section{Estimation}
\label{sec:estimation}

This section presents the two main estimation procedures for this model: first, we provide a brief discussion of estimation using Generalized Method of Moments (GMM), and then we provide a more elaborate description of the MLE estimator.
We will use notation similar to prior literature: $\Theta = \{\theta, \beta, \gamma\}$ represents all linear and non-linear parameters in the model. 
The parameters $(\beta,\gamma)$ enter the mean utility and cost equations linearly, and we refer to them as the linear parameters.
We refer to the remaining parameters $\theta = \{\alpha, \sigma_{\beta}\}$ as the non-linear parameters. 

\subsection{Generalized Method of Moments} 

For GMM estimation, the supply and demand residuals are used to form the moment conditions using valid instruments.
Identification relies on finding such valid instruments, which typically arise from exclusion restrictions, i.e., observables that enter only one of the demand and supply equations.
For a given set of parameters $\Theta$, Equations~\ref{eq:demand_reg} and \ref{eq:supply_reg} imply the following residuals:
\begin{align}
  \label{eq:GMM_residuals}
    \vect{u}_t(\Theta)  & = \vect{c}_t(\theta) -  \vect{w}_{t} \vect{\gamma},\\
    \vect{\xi}_{t}(\Theta) & = \vect{\delta}_t(\theta) - \vect{x}_t \vect{\beta} + \alpha \vect{p}_t.
\end{align}
As \cite{berry1994estimating} shows, the share equation~\ref{eq:share_int} is invertible in the sense that given the data $\{\mat{x}_{t}, \vect{p}_{t}, \vect{s}_{t}\}$ and non-linear parameters $\theta$, there is a unique value of the mean utilities $\delta_{jt}$ that rationalize the observed market shares.
We denote this by $\vect{\delta}_t(\theta)$.
Then as seen in the first equality of Equation \ref{eq:supply_reg_2}, the implied costs can be recovered from the prices, ownership matrix and the Jacobian of the shares with respect to prices.  
Notice that these also depend only on the non-linear parameters, and we write the implied costs as a function $\vect{c}_t(\theta)$.

The GMM estimator requires demand-side instruments $Z_{jt}^d$ and supply-side instruments $Z_{jt}^s$ that satisfy the moment conditions $E[\xi_{jt}Z_{jt}^d] = 0$ and $E[u_{jt}Z_{jt}^s] = 0$.
Prior work suggests various ways to construct these instruments~\citep{berry1995automobile,gandhi2019measuring,conlon2020best}. 
Given instruments, one can form the empirical analogs of the moment conditions:
\begin{equation}
    \label{eq:GMM_moments}
    \hat{g}(\theta) = \frac{1}{N} \sum_{t}\sum_{j \in \mathcal{J}_t}
    \begin{bmatrix} 
      u_{jt}(\theta) Z_{jt}^{s}\\
      \xi_{jt}(\theta) Z_{jt}^{d}
    \end{bmatrix} = 0,
\end{equation}
and proceed to minimize the GMM objective:
\begin{equation}
\label{eq:GMM_objective}
    \hat{\theta} = \argmin_{\theta} \hat{g}(\theta)\hat{\Psi}\hat{g}(\theta),
\end{equation}
where $\Psi$ is a positive definite weighting matrix.
In most cases, GMM estimation proceeds in two steps.
In the first step, $\Psi$ is typically set to either the identity matrix, or two-stage least squares (2SLS) weighting matrix.\footnote{In our simulations and replication exercise, we initialize $\Psi$ using the 2SLS weighting matrix as in~\cite{conlon2020best}.} In the second step $\Psi$ is replaced with a heteroscedasticity robust weighting matrix that uses the first-step estimates. More details on this estimation routine can be found in~BLP, AGS, and CG.\footnote{CG succinctly outlines the estimation procedure in their  Algorithm 1.}

\paragraph{Optimization}
We follow common practice and structure the optimization problem as an outer loop that searches
over non-linear parameters, and an inner loop that recovers the optimal linear parameters given the  non-linear parameters from the outer loop.
As noted above, for any guess of the non-linear parameters $\theta$ we can generate the implied mean utilities $\vect{\delta}_t(\theta)$ and marginal costs $\vect{c}_t(\theta)$.
We may then concentrate out the linear parameters $(\beta,\gamma)$ by setting them to the values that minimize the GMM objective for that $\theta$, a process referred to as the ``inner loop''.
The inner loop has an analytic solution, using an IV-GMM estimator, as described in~CG among other references.  

\paragraph{Standard Errors}
We use the standard formula for heteroskedasticity robust standard errors in GMM.
In practice the estimated standard errors depend on how one estimates the gradient of the GMM objective: either by finite differences, automatic differentiation, or using an analytic expression for the gradient. We primarily rely on automatic differentiation.

\subsection{Maximum Likelihood Estimation} 

In this section, we outline our maximum likelihood estimation (MLE) procedure.  
In this context MLE requires a parametric specification of the joint distribution of the error terms, $\xi$ and $u$. 
If this assumption holds, 
MLE is the most statistically efficient estimation procedure available.   
The identification requirements for MLE are otherwise the same as for GMM: the researcher should have access to a cost shifter to identify the mean price coefficient, and choice set variation to identify the distribution of random coefficients.\footnote{Instead of a cost shifter, one may instead impose that the cost and demand errors are independent, in which case the cost residual itself is a valid instrument for price \citep{mackeymiller21}.}  
Thus MLE requires demand and supply-side instruments just as GMM does, but doesn't require the explicit construction of moment conditions.
Since MLE can be thought of as GMM with optimal instruments (the first order conditions of the likelihood in the parameters), one can think of MLE as automatically picking the best instruments. This is distinct from the optimal GMM instruments of \cite{chamberlain1987asymptotic}, which are only semi-parametrically optimal, since that procedure makes no distributional assumptions on the distribution of the residuals.\footnote{In simulations, CG find that using the \citet{chamberlain1987asymptotic} instruments generally performs better than other approaches for picking instruments, but the performance gains they document from doing so are small compared to those we find from the use of MLE.}


The likelihood will depend on the joint density of the observed and endogenous variables in the model: shares ($\vect{s}_t$) and prices ($\vect{p}_t$). As mentioned previously, to estimate the likelihood, we make distributional assumptions. Particularly, we assume that the residuals of the demand and supply unobserved characteristics are i.i.d. (homoskedasticity of the residuals) and follow a joint normal distribution:
\begin{equation}
\label{eq:mle_structural_dist}
\begin{bmatrix}
  \xi_{jt} \\ 
  u_{jt} 
\end{bmatrix}  
\sim N(0, \Sigma) \; \text{, where }
\Sigma = \begin{bmatrix} 
  \sigma_{\xi}^2 \; \sigma_{\xi, u}\\
  \sigma_{\xi, u} \;  \sigma_{u}^2
\end{bmatrix}.
\end{equation}

The conditional density of interest is $f(\vect{s}_{t},  \vect{p}_{t}| \vect{x}_{t}, \vect{w}_{t} ;\vect{\Theta})$, where $f(\cdot)$ is probability density function. 
We can re-write this density by using multi-variate change of variables and the Inverse Function Theorem:
\begin{equation}
  \label{eq:mle_density}
  f(\vect{s}_{t}, \vect{p}_{t}| \vect{x}_{t}, \vect{w}_{t} ;\vect{\Theta}) = f(\vect{\delta}_{t},  \vect{c}_{t} | \vect{x}_{t}, \vect{w}_t;\vect{\Theta}) \; \times \; \left|  \mat{J}(\vect{s}_{t}, \vect{p}_{t}, \vect{x}_{t}, \vect{w}_{t};\vect{\theta})^{-1} \right|
\end{equation}
where $|\cdot|$ is the determinant and $\mat{J}(\vect{s}_{t},  \vect{p}_{t}| \vect{x}_{t}, \vect{w}_{t};\theta)$ are the full derivatives of shares and prices with respect to mean utility and costs, a $2N_t \times 2N_t $ matrix:
\begin{equation}
  \label{eq:jac_expansion}
           \mat{J}(\vect{s}_{t},  \vect{p}_{t}| \vect{x}_{t}, \vect{w}_{t};\vect{\theta}) = \begin{bmatrix} \frac{\mat{d s}_{t}}{\mat{d \delta}_{t}} \;  \frac{\mat{d s}_{t}}{\mat{d c}_{t}}\\
                          \frac{\mat{d p}_{t}}{\mat{d \delta}_{t}} \;  \frac{\mat{d p}_{t}}{\mat{d c}_{t}}\\
          \end{bmatrix}.
\end{equation}
We derive the Jacobian in  Appendix \ref{sec:jacobian}.
Notice that 
it doesn't  depend on the parameters $\{\vect{\beta}, \vect{\gamma}, \mat{\Sigma}\}$. The density of $f(\vect{\delta}_{t},  \vect{c}_{t} | \vect{x}_{t}, \vect{w}_t;\vect{\Theta})$ can be written as a result of the structure of the model (particularly equations \ref{eq:demand_reg} and \ref{eq:supply_reg_2}) and the distributional assumptions in equation \ref{eq:mle_structural_dist}:
\begin{equation}
  \label{eq:delta_c_density}
    f(\delta_{jt},  c_{jt} | \vect{x}_{jt}, \vect{w}_{jt};\vect{\Theta}, \mat{\Sigma}) \sim N(
    \mu(\vect{x}_{jt}, \vect{w}_{jt};\vect{\Theta}), \mat{\Sigma}),
\end{equation}
where $\mu$ is a vector-valued function (returning a two-vector), and $\Sigma$ is a $2 \times 2$ matrix.
For ease of notation, we denote the data in market $t$ as $\mat{\mathcal{D}}_t$ and the data across all markets as $\mat{\mathcal{D}}$. The log-likelihood can be written by expanding out the joint normal distribution for a market $t$ and taking the natural logarithm of all the terms in Equation \ref{eq:mle_density}:
\begin{align}
    \ell_t(\vect{\Theta}, \mat{\Sigma}; \mat{\mathcal{D}}_t) = \nonumber
    & -\frac{N_t}{2}\log(|\mat{\Sigma}|) \\ \nonumber
    & -\frac{1}{2} \sum_{j \in \mathcal{J}_t}\left( ([\delta_{jt} \;  c_{jt}] - \mu(\vect{x}_{jt}, \vect{w}_{jt};\vect{\Theta})^T)\mat{\Sigma}^{-1} (\begin{bmatrix}\delta_{jt} \\  c_{jt}\end{bmatrix}- \mu(\vect{x}_{jt}, \vect{w}_{jt};\vect{\Theta)}) \right)\\ \nonumber
    & -\log(|\mat{J}( \mat{\mathcal{D}}_t;\vect{\theta})|).
\end{align}
Write $\begin{bmatrix}\xi _{jt}(\Theta) \\  u_{jt}(\Theta) \end{bmatrix} =  \begin{bmatrix}\delta_{jt}  \\  c_{jt}\end{bmatrix}- \mu(\vect{x}_{jt}, \vect{w}_{jt};\vect{\Theta})$, substitute into the likelihood and sum across markets to get the aggregate log-likelihood:

\begin{align}
  \label{eq:ell}
  \ell(\vect{\Theta}, \mat{\Sigma}; \mat{\mathcal{D}}) &= \sum_t \ell_t(\vect{\Theta}, \mat{\Sigma}; \mat{\mathcal{D}}_t)\\ \nonumber
  &= \sum_t\left( - \frac{N_t}{2}\log(|\mat{\Sigma}|)
    -\frac{1}{2} \sum_{j \in \mathcal{J}_t} [\xi_{jt} \;  u_{jt}]^T\mat{\Sigma}^{-1} \begin{bmatrix}\xi_{jt} \;  u_{jt}\end{bmatrix} - \log(|\mat{J}( \mat{\mathcal{D}}_t;\vect{\theta})|)\right).
\end{align}
We will refer to the  log-likelihood given above as the \textit{unconcentrated} version of the log-likelihood. 


\paragraph{Optimization}
To simplify optimization, we will concentrate out a number of parameters.
First consider the variance-covariance matrix $\Sigma$.
Fixing $\Theta$ and therefore the  residuals $\{\xi_{jt}(\Theta),u_{jt}(\Theta)\}$, the likelihood maximizing choice of $\Sigma$ is the implied  variance-covariance matrix of the residuals.
This is a function of the parameters, which we denote by  $\Sigma^*(\Theta)$.  
As shown in Appendix \ref{sec:part_2_simp}, making this substitution simplifies the middle term of the log-likelihood in eq~\ref{eq:ell} to a constant, resulting in the following likelihood:
\begin{equation}
\ell(\vect{\Theta}, \mat{\Sigma}; \mat{\mathcal{D}}) =  - \frac{N}{2}\log(|\mat{\Sigma}^*(\Theta)|) - \sum_t \log(|\mat{J}( \mat{\mathcal{D}}_t;\vect{\theta})|).
\end{equation}
Notice that the linear parameters do not enter the Jacobian, only the implied variance-covariance matrix.  
Thus similar to the GMM procedure, we can concentrate out the linear  parameters by solving an inner loop problem of the form:
\begin{equation}
    \label{eq:covariance_obj}
    \vect{\gamma}^*(\theta), \vect{\beta}^*(\theta) = \argmin_{\gamma, \beta}\log(|\mat{\Sigma^*(\theta, \beta, \gamma)}|).
\end{equation}
Now recall that the implied variance covariance-matrix $\Sigma^*(\theta, \beta, \gamma)$ is a  $2 \times 2$ matrix, with a determinant equal to $\sigma^2_{\xi}(\theta, \beta, \gamma)\sigma^2_{u}(\theta, \beta, \gamma) - (\sigma_{\xi,u}(\theta, \beta, \gamma))^2$.
In the absence of a covariance term, minimizing this objective would be equivalent to minimizing the demand and cost residuals separately.
This could be done by separate OLS regressions.
But the presence of a covariance term complicates things, leading to a non-linear objective with no closed form solution.\footnote{In the GMM estimation procedure, OLS also fails unless there is no covariance in the errors.  But there IV-GMM delivers a closed-form solution for $(\beta,\gamma)$.}
In Appendix~\ref{sec:variance_estimator} we show that $\beta$ and $\gamma$ can be solved in closed form  holding the other parameter fixed and so we estimate them jointly by alternating least squares until convergence.

This yields a concentrated likelihood function that depends only on the non-linear parameters, to be maximized in an outer loop:
\begin{equation}
  \label{eq:likelihood}
\ell(\vect{\theta};  \mat{\mathcal{D}}) = 
- \underbrace{\vphantom{\sum_{t}}\frac{N}{2}\log(|\mat{\Sigma}^*(\theta)|)}_\text{Part I} 
- \underbrace{\sum_{t}\log(|\mat{J( \vect{\theta})}|)}_\text{Part II},
\end{equation}
where we will call part I of the likelihood the \textit{covariance term} and part II the \textit{Jacobian term}. 

\paragraph{Standard Errors}
To compute standard errors, we use the inverse of the Fisher information matrix, which is given in Appendix \ref{sec:std_err}. For accurate estimation of the standard errors, we use the Fisher information matrix of the unconcentrated likelihood, to allow the uncertainty in the inner loop estimation procedures to be accounted for.
 
\paragraph{Computational Details and Algorithm}
The optimization of the likelihood initially follows the GMM procedure in inverting from market shares to mean utilities and marginal cost vectors $\{\mat{\delta},\mat{c}\}$. In order to invert from market shares to mean utilities, we use the SQUAREM method proposed by \cite{varadhan2008simple} and recommended by CG. The marginal cost is calculated from the Bertrand-Nash equations detailed in the Model section.

The rest of the optimization is best understood when thinking about the \textit{covariance} term and the \textit{Jacobian} term of the log-likelihood in equation \ref{eq:likelihood} separately. The covariance term is evaluated by concentrating out the linear terms using the objective in equation \ref{eq:covariance_obj}.  

The terms entering the Jacobian terms depend on partial hessians of the shares with respect to price.
They are derived in detail in Appendix \ref{sec:hessian}.
Note that these Hessians are the computational bottleneck of the estimation routine, having dimensions of size $(T, N_i, max(N_t), max(N_t), max(F_t))$\footnote{The $max(N_t)$ is a result of how we programmed markets. Essentially, in order to have all markets concatenated in a matrix or tensor, the max number of products is used. Additionally, the $max(F_t)$ dimension is a result of the particular hessian objects having a dimension that has to only do with products that are within the same firm.}. 

Given data, Algorithm \ref{alg:main} outlines the procedure of the estimation objective. The code developed for this application uses the PyTorch package. PyTorch provides a high-performance library for optimizing both the GMM and MLE objectives.\footnote{Two reasons for using PyTorch are: (1) utilizing the efficient tensor and matrix operations, (2) utilizing automatic differentiation for optimization.} 

\begin{algorithm}
\caption{MLE estimation of BLP}\label{alg:main}
\begin{algorithmic}
\Require $\{\theta, data\}$
\State $\xrightarrow{}$ Calculate $\delta(\theta | data)$ using the SQUAREM algorithm.
\State  $\xrightarrow{}$ Calculate $c(\theta|data)$ using Equation \ref{eq:supply_foc}. \State  $\xrightarrow{}$ Calculate $\{\beta^*, \gamma^*\}$ using the ALS process described in appendix section \ref{sec:variance_estimator}.
\State  $\xrightarrow{}$ Calculate $\ell_{I}(\theta, \beta^*, \gamma^*|data)$, part I of the log-likelihood equation \ref{eq:likelihood}.
\State $\xrightarrow{}$ Calculate $\ell_{II}(\theta | data)$, part II of the log-likelihood equation \ref{eq:likelihood} as described in appendix sections \ref{sec:jacobian} and \ref{sec:hessian}.

\State \textbf{return } $\ell = \ell_{I} + \ell_{II}$ 
\end{algorithmic}
\end{algorithm}

\section{Simulations}
\label{sec:simulation}

In this section, we evaluate the performance of GMM and MLE on simulated data, where the ground truth is known.
Our simulation procedure is based on \cite{conlon2020best} and \cite{armstrong2016large}. 




For each simulation, we set the number of markets to $T = 20$.
The number of firms $F_t$ in each market is randomly chosen from the set $\{2, 5, 10\}$. 
The number of products for each firm $\mathcal{J}_{ft}$ is also chosen randomly from $\{3, 4, 5\}$. 
The structural error terms are given by:
\[
  \begin{bmatrix}\xi_j \\ u_j  \end{bmatrix} \sim N(0, \Sigma) \text{ where }
  \Sigma = \begin{bmatrix} 0.2 \; 0 \\ 0 \; 0.2 \end{bmatrix}.
\]
The linear demand characteristics are $[1, x_{jt}, p_{jt}]$ and the linear supply characteristics are $[1, x_{jt}, w_{jt}]$.
Both exogenous characteristics ($x_{jt}, w_{jt}$) are drawn from a standard uniform distribution.
We allow for two random coefficients on the demand side. 
The first is on the demand characteristic $x_{jt}$, where $\tilde{\beta}_{i} \sim N(0, \sigma_{\beta}^2)$ and $\sigma_{\beta} = 3$. 
The second one is on price, where $\tilde{\alpha}_{i} \sim N(0, \sigma_{\alpha}^2)$ and $\sigma_{\alpha}=0.2$. 
To compute the values of the endogenous variables ($p_{jt}, s_{jt}$) we use the~\cite{morrow2011fixed} method as described in~\cite{conlon2020best}.
We use Gauss-Hermite quadrature with a product rule to integrate over the 2-dimensional distribution of random coefficients. We provide details in Appendix~\ref{sec:quadrature}.
We set the ground-truth values for the demand-side parameters to $[\beta_0, \beta_x, \alpha] = [−7, 6, −1]$, and the supply-side parameters to $[\gamma_0, \gamma_x, \gamma_w] = [2, 1, 0.2]$.
To ensure reproducibility, we fix a random seed for each simulation.

We refer to the setup above as ``No covariance'', and also consider the following variations:
\begin{itemize}
    \item \textbf{Low covariance:} We set the variance-covariance matrix to $\Sigma = \begin{bmatrix} 0.2 \; 0.1 \\ 0.1 \; 0.2 \end{bmatrix}$.
    \item \textbf{High covariance:} We set the variance-covariance matrix to $\Sigma = \begin{bmatrix} 0.3 \; 0.2 \\ 0.2 \; 0.3 \end{bmatrix}$.
    \item \textbf{Laplace, no and low covariance:}
    The errors are generated so that the marginals are Laplace instead of normal. A Gaussian copula is used to generate the draws in the low covariance case, where $\Sigma$ takes the form above.
    \item \textbf{Supply Misspecification:} The supply side functional form is misspecified. Marginal costs are linear in characteristics as per Equation \ref{eq:supply}, but the estimation assumes log-linearity $ln(c_{jt}) = w_{jt} \gamma + u_{jt}$.
    \item \textbf{Ownership Misspecification:} The ownership matrix is misspecified, in that we set $O$ to be the identity matrix (i.e. each product is assumed to be owned by a single firm). The truth is that each firm produces 3, 4 or 5 products. 
\end{itemize}

For GMM estimation, we construct local differentiation instruments following \cite{gandhi2019measuring}. 
We also create a predicted price instrument, by regressing price on valid instruments \citep{reynaert2014improving}.

\subsection{Results}

We compare MLE and GMM for each of the above scenarios on three metrics: parameter estimates, standard errors, and own-price elasticities. 
For each scenario, we run 1000 simulations, with a different random seed for each simulation\footnote{ Each random seed is also randomly chosen. The random seed is assigned for reproducibility.}. 
For each simulation, optimization starts at a random initial guess which is drawn uniformly within a 50\% range of the true values. 
Optimization is repeated 3 times with different starting values, after which the parameter estimates are chosen based on the smallest objective value. 

To validate that we have correctly coded up the scenarios in CG, as well as the our particular implementation of GMM, we compare our estimates to those produced by PyBLP. 
The parameter estimates are in most cases indistinguishable.
In addition, the mean bias and mean absolute bias we report in our GMM results closely match those in Table 5 of CG.

The scenarios that we run simulations for can be viewed in three different categories. First, the \textit{correctly specified} scenarios, where both the MLE and the GMM estimators are correctly specified. Second, the \textit{MLE misspecified} scenarios, which are the Laplace scenarios (GMM is correctly specified in this case). Last, there are the \textit{fully misspecified} scenarios, where both MLE and GMM are misspecified, these are the Supply Misspecification and Ownership Misspecification scenarios. 

\begin{figure}[t!]
    \centering
    \subfloat[No covariance scenario]{\label{b1}\includegraphics[width=\textwidth]{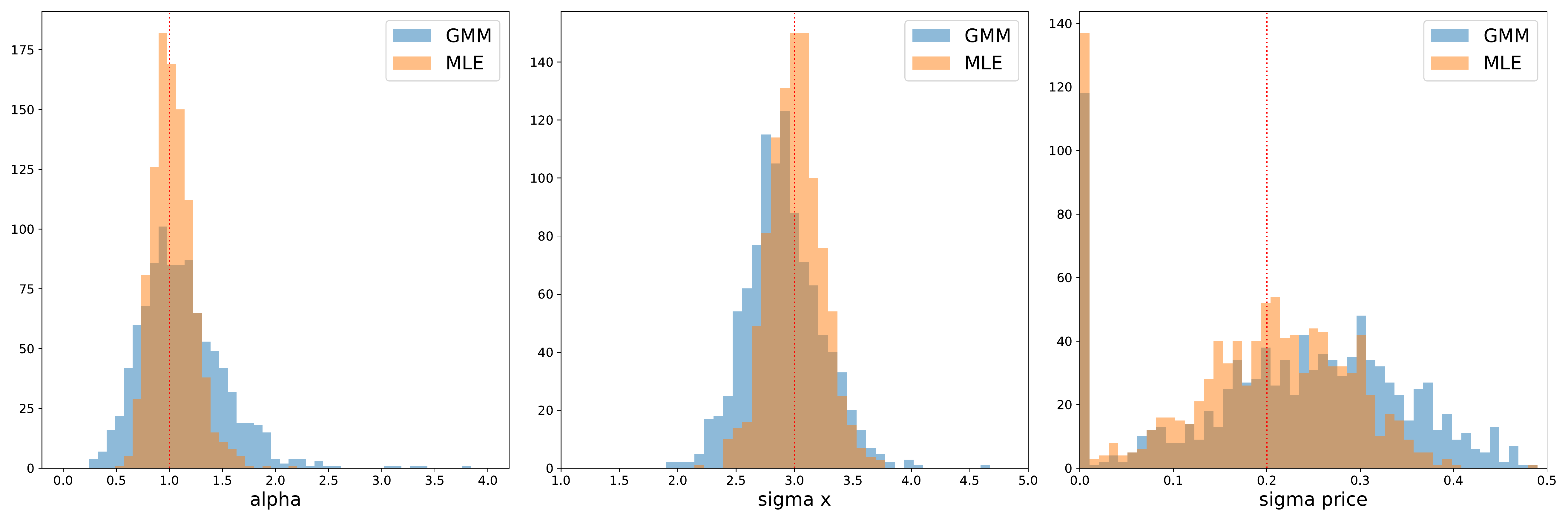}}\par 
    \subfloat[Low covariance scenario]{\label{c1}\includegraphics[width=\textwidth]{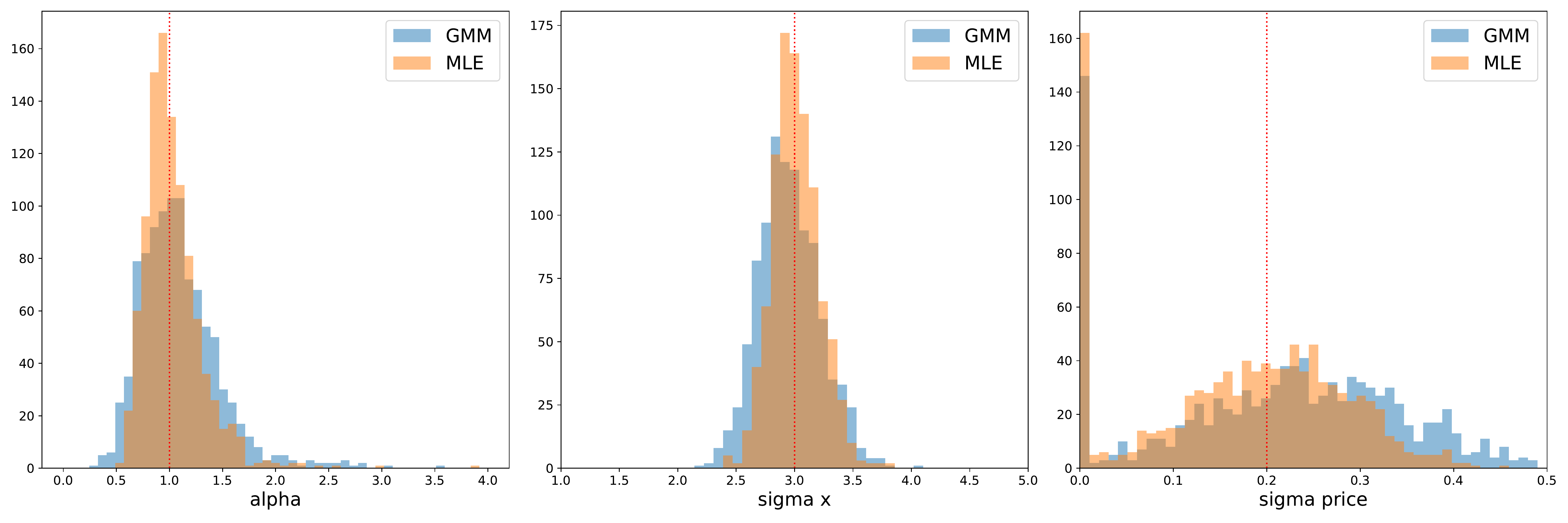}}
\caption{These figures plot the histograms of parameters across the 1000 simulation estimates for the (a) no covariance scenario and the (b) low covariance scenario. The dashed vertical line in each sub-figure is placed at the true parameter value. The results show the overall distribution of parameter estimates.}
\label{fig:param_hist}
\end{figure}

\paragraph{Mean Bias and RMSE.}
We begin by summarizing the bias and root mean squared error (RMSE) of the estimators, in two Tables: (a) the RMSE of the estimators can be found in Table \ref{tab:rmse_summary} and (b) the mean bias can be found in Table \ref{tab:mean_bias_summary}. The results show that in general the MLE estimates  perform somewhat better when it comes to mean bias, and clearly outperform the GMM estimates when looking at the RMSE values. Another way of putting this is for a particular realization of the data, MLE estimates are generally closer to the true values. A visualization of this can be seen in Figure \ref{fig:param_hist}, where we see the histogram of the parameter estimates of each of the three parameters for the no covariance and low covariance scenarios.

For RMSE, MLE clearly outperforms the GMM for all the correctly specified and MLE misspecified scenarios. 
In the fully misspecified scenarios, GMM performs better when there is a supply side misspecification and MLE performs better when the ownership matrix is misspecified. 




\paragraph{Standard Errors and Coverage.}
Standard errors and confidence intervals are fundamental for inference.
These results are summarized in two tables: (a) the mean standard error for a scenario in Table \ref{tab:mean_se} and (b) the coverage percentile of the true values for the 95\% confidence intervals in Table \ref{tab:percent_in_ci}.\footnote{For the standard errors of the $\sigma_{price}$ parameter in the GMM procedure, we find that about $10\%$ of the simulations have numerically infeasible values (standard error values ranging from 1000 to 1e13). To validate that this is not an error in our standard error computations, we take a handful of simulations with such issues and run the pyBLP package provided by CG. The outputs were $NaN$ for the corresponding standard error values. It therefore appears that this problem is not specific to our code, but to the GMM procedure implemented here and in pyBLP itself. We drop these cases in the relevant columns of Tables~\ref{tab:mean_se} and~\ref{tab:percent_in_ci}; if they were included the performance of GMM would be much worse.} The coverage percentile represents the percent of simulations where:
\begin{equation}
    \theta_{true} \in [\hat{\theta}-1.96\hat{\sigma}_{\theta}/\sqrt{n},  \hat{\theta}+1.96\hat{\sigma}_{\theta}/\sqrt{n}].
\end{equation}

The mean standard errors for MLE are tighter under all correctly specified scenarios as well as the Laplace  scenarios (the one exception to this is the mean standard error of $\alpha$ in the high covariance scenario). 
Table \ref{tab:percent_in_ci} shows the coverage of each estimator for each parameter, for a nominal 95\%.
In general GMM tends to badly under-cover, with true coverage of  between 80-90\% for the mean price parameter, 90-95\% for the standard deviation of the random coefficient on $x$ and 70-80\% for the standard deviation of the random coefficient on price.
By contrast for MLE those coverages are respectively 88-95\%, 90-95\% and 90-95\%, so that the undercoverage problem is less severe.  
The sole exception to this general pattern is the case where the cost structure is misspecified, where MLE performs terribly.


\begin{figure}[t!]
    \centering
    \subfloat[No covariance scenario]{\label{b}\includegraphics[width=.7\textwidth]{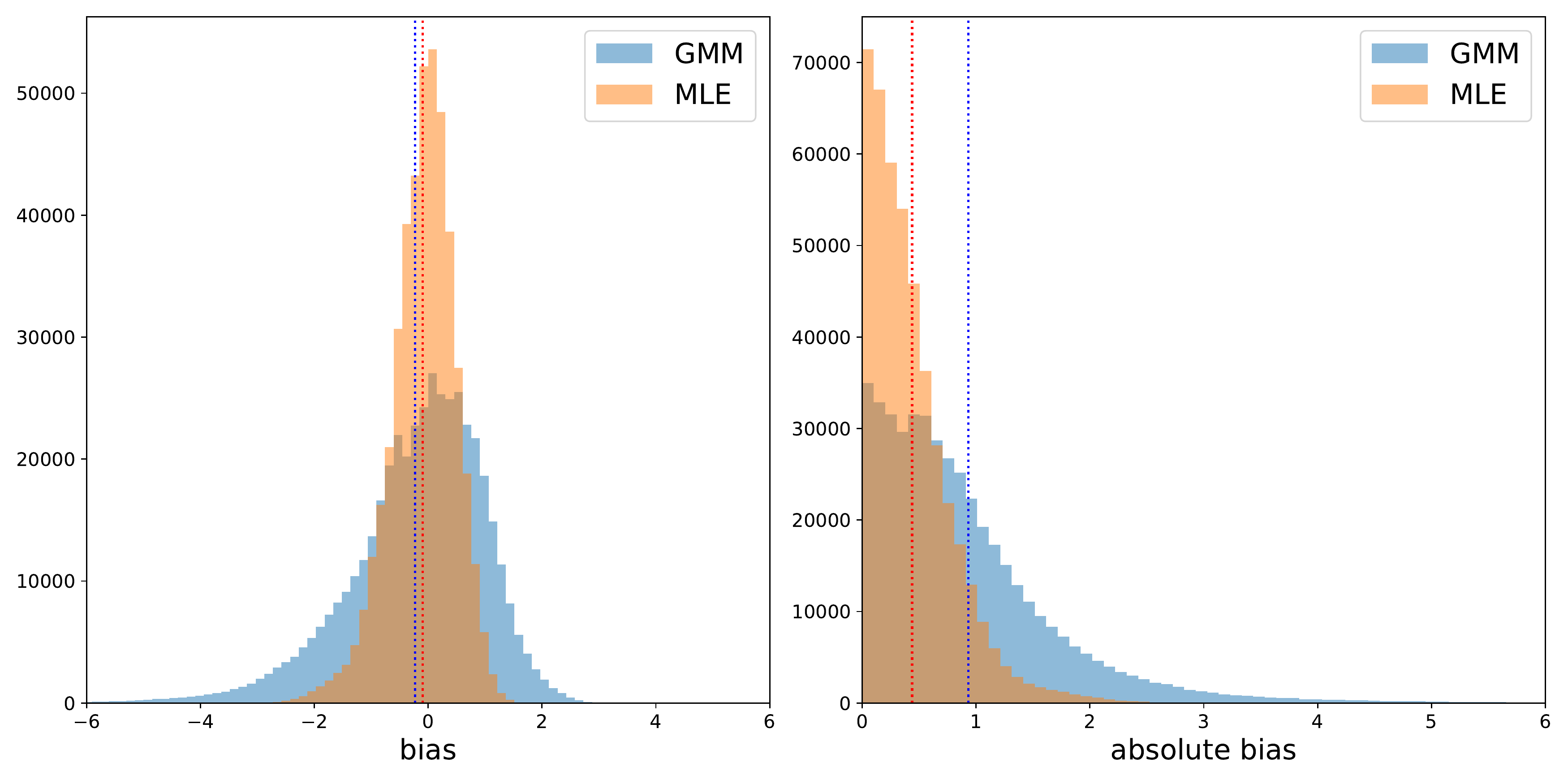}}\par 
    \subfloat[Low covariance scenario]{\label{c}\includegraphics[width=.7\textwidth]{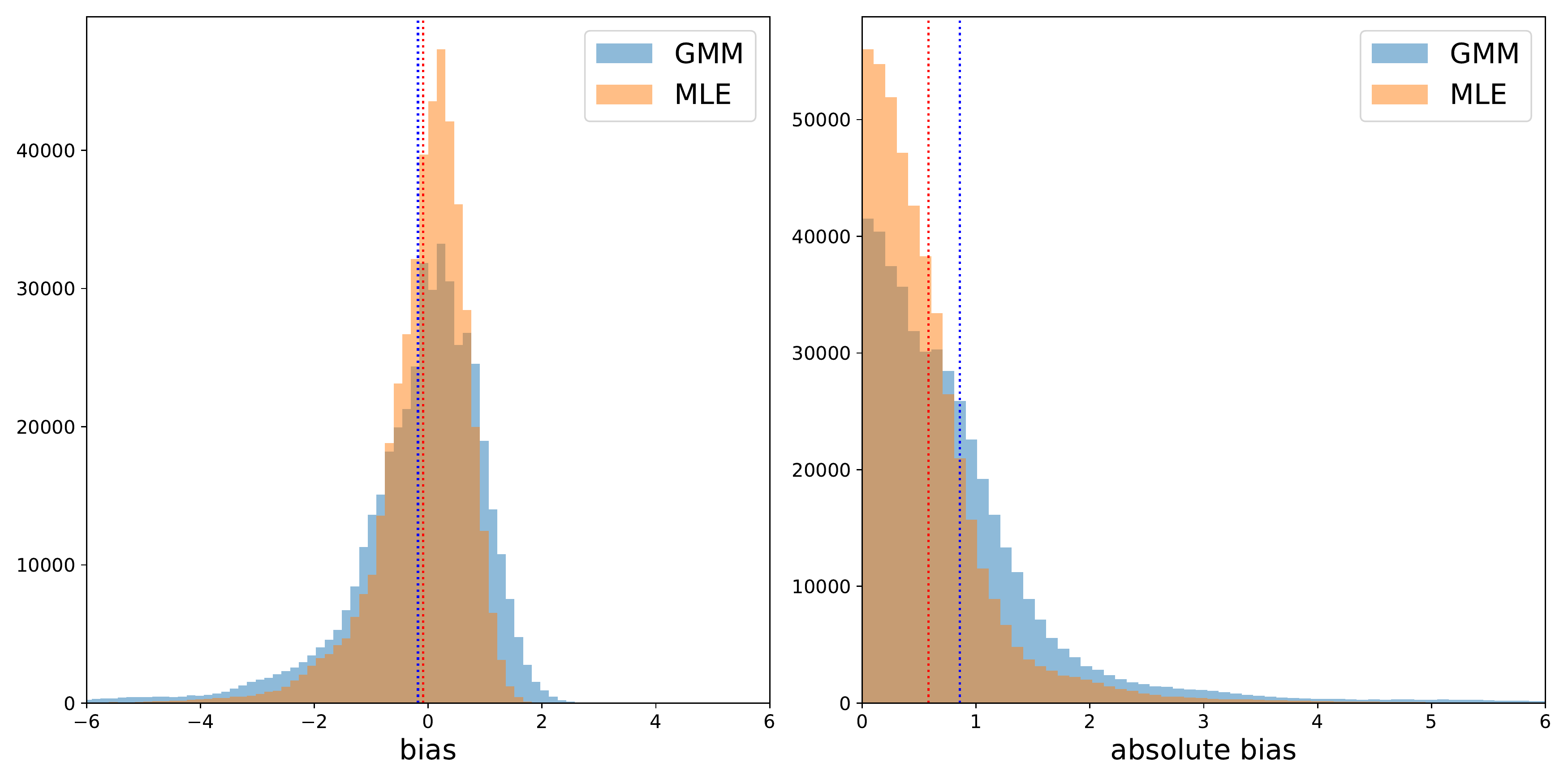}}
\caption{This figure has a set of histograms summarizing the bias (left) and absolute bias (right) of product-level own-price elasticities in the  (a) no covariance and (b) low covariance scenarios. Within each subfigure, the vertical dashed lines are the average bias and average absolute bias for GMM (blue) and MLE (red).}
\label{fig:own_price}
\end{figure}

\paragraph{Own-price elasticities} Demand systems are often used to generate own-price elasticities.
Given the interest in these statistics, we 
examine what the mean bias and absolute mean bias of product-level own-price elasticities are for a given set of estimated parameters. Figure \ref{fig:own_price} shows this for the no covariance and low covariance scenarios. The results support the conclusion from the previous sections, the MLE produces tighter results in most of our scenarios.

\begin{figure}[t!]
  \centering 
  \includegraphics[width=\textwidth]{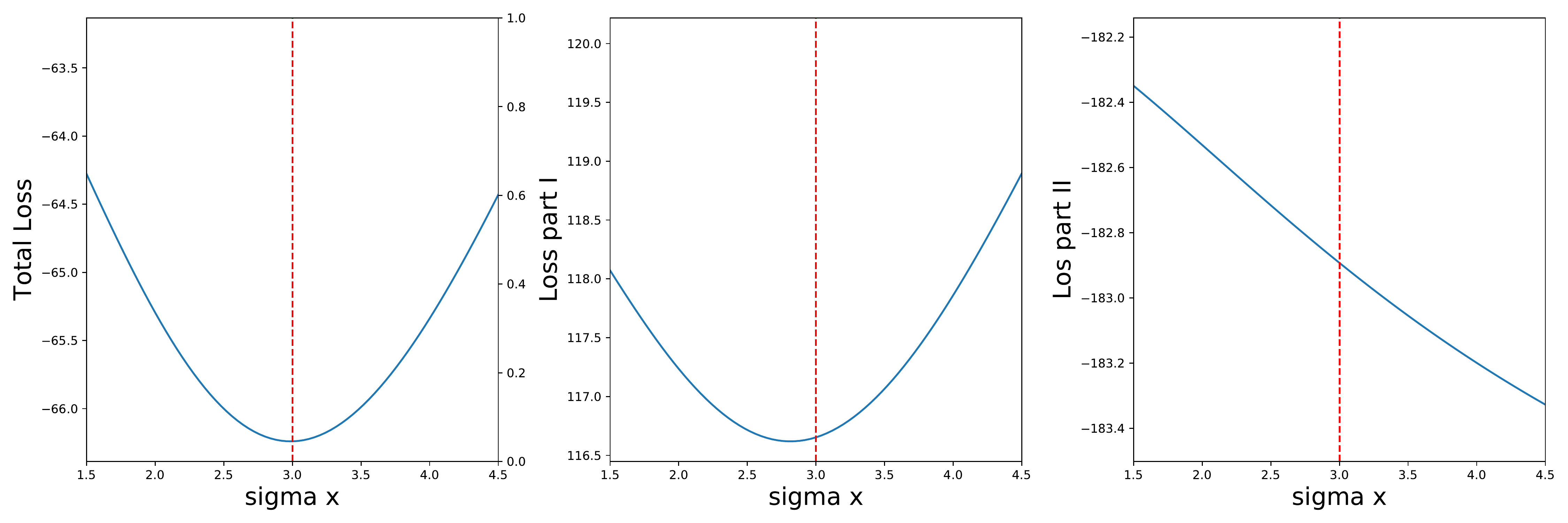}
  \caption{This figure plots the likelihood and its components for different values of $\sigma_x$. Each subplot has the same x-axis, which is a range of $\sigma_x$ around the true value of 3, while $\alpha$ and $\sigma_{price}$ stay fixed at the true value of 1. The figure on the left shows the full likelihood detailed in equation \ref{eq:likelihood}, with the dashed green line highlighting the GMM loss. The middle figure is the covariance term of the likelihood, and the figure on the right is the Jacobian term of the likelihood.  
  Finally, the vertical dashed red line is the true value of sigma.}\label{fig:likelihood_decomp}
\end{figure}

\paragraph{Jacobian.}
One notable feature of MLE is the presence of a Jacobian in the likelihood.
This accounts for the non-linear mapping from marginal costs and mean utilities to the (endogenous) prices and quantities.
One might think that minimizing the covariance term in the likelihood would be enough to ensure good performance, since doing so in some sense minimizes the variance of the demand and cost residuals. 
We investigate the importance of the Jacobian term in the likelihood here.

 
Figure \ref{fig:likelihood_decomp} plots the likelihood decomposition into the covariance and Jacobian terms separately, where we vary the $\sigma_x$ parameter at the true $\alpha=1$ value (for a random simulation). The covariance term is minimized below the truth, and the Jacobian term is necessary to correct the bias in the covariance term and arrive at the correct solution. 
The Jacobian thus plays an important role, in some sense regularizing the estimator by penalizing parameters under which the derivatives of the endogenous variables in the exogenous variables are big (i.e. parameters at which the predictions are unstable).  

\section{Empirical exercise: replicating BLP'95}
\label{sec:empirics}

In this section, we describe our replication of BLP'95 using MLE.
We use the same data as \cite{berry1995automobile}, provided by \cite{andrews2017measuring}. 
The data contains US automobile data from 1971 to 1990. The dataset also includes 5 product characteristics, horse power to weight ratio (hpwt), if a car has air conditioning (air), miles per dollar of gasoline (mpd), width times length (space), and miles per gallon (mpg).

We start by replicating the GMM estimation procedure. We follow the procedure outlined in AGS, which is based off of BLP. This includes dropping certain instruments due to collinearity as well as starting the estimation procedure from the estimated parameters in BLP. As in BLP, we have 2217 model/years with 997 distinct models. The instruments that we use are BLP sums of characteristic instruments. With this setup, the local GMM estimation produces estimates that are almost identical to the AGS parameters. 

Next, we estimate the parameters using MLE. For the MLE setup, we run the procedure on the same dataset. Our integration approach follows AGS by using the same random draws and importance sampling. One change we make is to double the number of draws to ensure symmetry. More specifically, we take the set of draws, multiply the vector associated with air conditioning with negative one, and concatenate these to the original draws. This leaves us with two times the total draws.\footnote{When estimating the model by MLE using the AGS draws, the parameter estimates were mostly similar, but 
$\sigma_{air}$ value was estimated to be 0.
This is a result of an artificial lack of symmetry around zero of the objective function in $\sigma_{air}$ due to sampling variance in the draws; doubling the number of draws fixes this problem.}


To get a sense of how reasonable our results are, we compare our estimates to existing benchmarks in the literature that use this dataset. These benchmarks are the estimates reported in BLP, AGS, and CG. These benchmarks differ in two main ways, the instruments selected and the choice of draws (and weights) used for integration. BLP use the sum of rival characteristics (BLP) instruments and integrate using importance sampling. AGS follows BLP closely in order to replicate their estimates. CG uses the \cite{chamberlain1987asymptotic} optimal instruments and quasi-Monte Carlo integration (10,000 scrambled Halton draws in each market).

The estimation results can be found in Table \ref{tab:blp95_param_se_estimates} with standard error comparisons in parentheses below the parameter estimates. The parameter estimates of MLE are closely aligned with the GMM estimates. In particular, the $\alpha$ parameter is closest to the CG best practice estimates. The sigma estimates tend to align with the BLP results, and generally indicate larger heterogeneity in the data than the CG results. The similarity of the parameter estimates is surprising since the effort put into calibrating the MLE estimator to the automobile data was minimal i.e. these results are ``out-of-the-box''. 
No calibration or instrument construction was necessary.
Standard error estimates of the MLE parameters are significantly tighter for MLE, regularly one quarter of the smallest GMM standard error. The standard error on $\alpha$ is 0.769, compared to a range of 5.6-11.7 for the GMM. This is nearly 10-fold reduction compared to GMM. Since we don't know the ground truth we cannot comment on whether the estimated standard errors imply confidence intervals with correct coverage, but the simulations above give us confidence that if the model is correctly specified the coverage will be approximately correct.  

\begin{figure}[t!]
  \centering 
  \includegraphics[width=.65\textwidth]{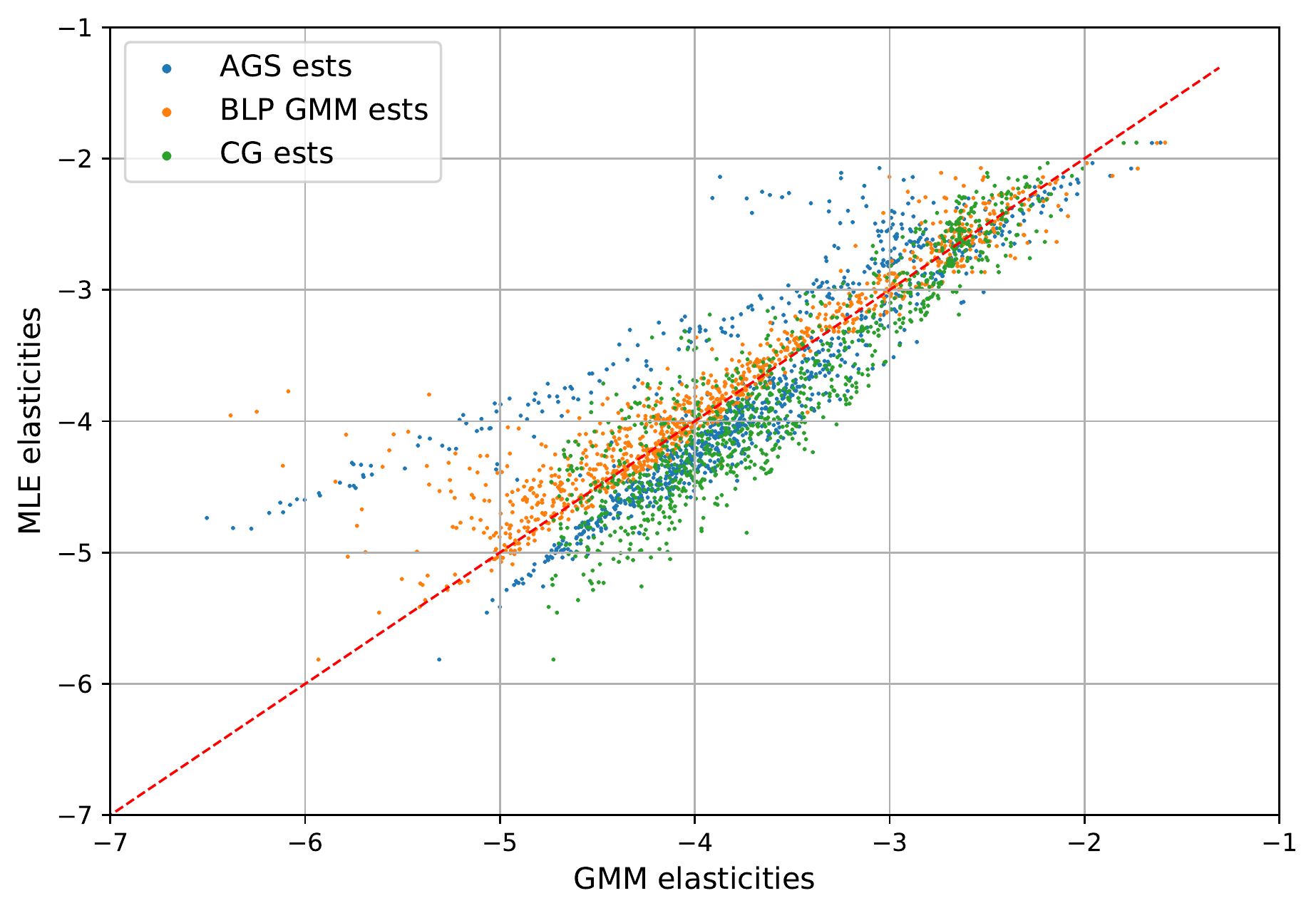}
  \caption{This figure plots the own-price elasticity comparisons between MLE and GMM parameters. Each dot represents a model own-price elasticity, with the different colors representing the three main GMM estimates we use as benchmarks in this paper.}\label{fig:own_blp95_gmm_mle}
\end{figure}

Finally, we look at the model level comparison of the own-price elasticities of the GMM and MLE estimators in Figure \ref{fig:own_blp95_gmm_mle}. 
We can see that there is a fairly large mass around the diagonal, indicating generally similar estimates. The most significant trend off of the diagonal is a section of the AGS estimates. This trend can be fully attributed to the large difference in the $\sigma_{air}$ estimates, where AGS estimates this parameter to be 4.2, compared to BLP, CG, and MLE estimating it to between 1.4-2. Since the air conditioning variable is binary, it leads to two separate trend lines, where the one above the diagonal representing models that have air conditioning.

Overall, we were able to use MLE on the BLP automobile data with relative ease and produce estimates that are similar to prior work but with tighter standard errors and without having to the construct valid instruments.

\section{Conclusion}
\label{sec:conclusion}

We develop and evaluate a maximum likelihood estimator for differentiated products demand. 
Unlike the traditionally used GMM estimator, the ML estimator requires stricter distributional assumptions, while offering the advantages of statistical efficiency and limiting choices required by the researcher to fit the model.
While we are not the first to propose an ML estimator, prior work has relied on reduced-form supply specifications, whereas our supply-side pricing is consistent with Bertrand-Nash pricing.

In the simulation section, we show that under correct specification, our method outperforms GMM as well as showing its robustness and sensitivity to different types of misspecification. Finally, we replicate BLP using the new estimation method, finding that the estimates are very similar by comparing the estimated values to other benchmarks. These results demonstrate that MLE can be a useful addition to the tools available to researchers estimating BLP.

\bibliography{References}

\begin{thebibliography}{20}
\providecommand{\natexlab}[1]{#1}
\providecommand{\url}[1]{\texttt{#1}}
\expandafter\ifx\csname urlstyle\endcsname\relax
  \providecommand{\doi}[1]{doi: #1}\else
  \providecommand{\doi}{doi: \begingroup \urlstyle{rm}\Url}\fi

\bibitem[Abaluck and Adams-Prassl(2021)]{abaluck2021consumers}
J.~Abaluck and A.~Adams-Prassl.
\newblock What do consumers consider before they choose? identification from
  asymmetric demand responses.
\newblock \emph{The Quarterly Journal of Economics}, 136\penalty0 (3):\penalty0
  1611--1663, 2021.

\bibitem[Andrews et~al.(2017)Andrews, Gentzkow, and
  Shapiro]{andrews2017measuring}
I.~Andrews, M.~Gentzkow, and J.~M. Shapiro.
\newblock Measuring the sensitivity of parameter estimates to estimation
  moments.
\newblock \emph{The Quarterly Journal of Economics}, 132\penalty0 (4):\penalty0
  1553--1592, 2017.

\bibitem[Armstrong(2016)]{armstrong2016large}
T.~B. Armstrong.
\newblock Large market asymptotics for differentiated product demand estimators
  with economic models of supply.
\newblock \emph{Econometrica}, 84\penalty0 (5):\penalty0 1961--1980, 2016.

\bibitem[Berry et~al.(1995)Berry, Levinsohn, and Pakes]{berry1995automobile}
S.~Berry, J.~Levinsohn, and A.~Pakes.
\newblock Automobile prices in market equilibrium.
\newblock \emph{Econometrica: Journal of the Econometric Society}, pages
  841--890, 1995.

\bibitem[Berry(1994)]{berry1994estimating}
S.~T. Berry.
\newblock Estimating discrete-choice models of product differentiation.
\newblock \emph{The RAND Journal of Economics}, pages 242--262, 1994.

\bibitem[Chamberlain(1987)]{chamberlain1987asymptotic}
G.~Chamberlain.
\newblock Asymptotic efficiency in estimation with conditional moment
  restrictions.
\newblock \emph{Journal of econometrics}, 34\penalty0 (3):\penalty0 305--334,
  1987.

\bibitem[Conlon and Gortmaker(2020)]{conlon2020best}
C.~Conlon and J.~Gortmaker.
\newblock Best practices for differentiated products demand estimation with
  pyblp.
\newblock \emph{The RAND Journal of Economics}, 51\penalty0 (4):\penalty0
  1108--1161, 2020.

\bibitem[Conlon(2013)]{conlon2013empirical}
C.~T. Conlon.
\newblock The empirical likelihood mpec approach to demand estimation.
\newblock \emph{Available at SSRN 2331548}, 2013.

\bibitem[Dub{\'e} et~al.(2012)Dub{\'e}, Fox, and Su]{dube2012improving}
J.-P. Dub{\'e}, J.~T. Fox, and C.-L. Su.
\newblock Improving the numerical performance of static and dynamic aggregate
  discrete choice random coefficients demand estimation.
\newblock \emph{Econometrica}, 80\penalty0 (5):\penalty0 2231--2267, 2012.

\bibitem[Gandhi and Houde(2019)]{gandhi2019measuring}
A.~Gandhi and J.-F. Houde.
\newblock Measuring substitution patterns in differentiated-products
  industries.
\newblock Technical report, National Bureau of Economic Research, 2019.

\bibitem[Honka et~al.(2017)Honka, Horta{\c{c}}su, and
  Vitorino]{honka2017advertising}
E.~Honka, A.~Horta{\c{c}}su, and M.~A. Vitorino.
\newblock Advertising, consumer awareness, and choice: Evidence from the us
  banking industry.
\newblock \emph{The RAND Journal of Economics}, 48\penalty0 (3):\penalty0
  611--646, 2017.

\bibitem[Jiang et~al.(2009)Jiang, Manchanda, and Rossi]{jiang2009bayesian}
R.~Jiang, P.~Manchanda, and P.~E. Rossi.
\newblock Bayesian analysis of random coefficient logit models using aggregate
  data.
\newblock \emph{Journal of Econometrics}, 149\penalty0 (2):\penalty0 136--148,
  2009.

\bibitem[MacKay and Miller(2021)]{mackeymiller21}
A.~MacKay and N.~Miller.
\newblock Estimating models of supply and demand: Instruments and covariance
  restrictions.
\newblock Technical report, Harvard University, 2021.

\bibitem[McFadden(1986)]{mcfadden86}
D.~McFadden.
\newblock The choice theory approach to market research.
\newblock \emph{Marketing science}, 5\penalty0 (4):\penalty0 275--297, 1986.

\bibitem[Morrow and Skerlos(2011)]{morrow2011fixed}
W.~R. Morrow and S.~J. Skerlos.
\newblock Fixed-point approaches to computing bertrand-nash equilibrium prices
  under mixed-logit demand.
\newblock \emph{Operations research}, 59\penalty0 (2):\penalty0 328--345, 2011.

\bibitem[Park and Gupta(2009)]{park2009simulated}
S.~Park and S.~Gupta.
\newblock Simulated maximum likelihood estimator for the random coefficient
  logit model using aggregate data.
\newblock \emph{Journal of Marketing Research}, 46\penalty0 (4):\penalty0
  531--542, 2009.

\bibitem[Paszke et~al.(2019)Paszke, Gross, Massa, Lerer, Bradbury, Chanan,
  Killeen, Lin, Gimelshein, Antiga, et~al.]{paszke2019pytorch}
A.~Paszke, S.~Gross, F.~Massa, A.~Lerer, J.~Bradbury, G.~Chanan, T.~Killeen,
  Z.~Lin, N.~Gimelshein, L.~Antiga, et~al.
\newblock Pytorch: An imperative style, high-performance deep learning library.
\newblock \emph{Advances in neural information processing systems},
  32:\penalty0 8026--8037, 2019.

\bibitem[Reynaert and Verboven(2014)]{reynaert2014improving}
M.~Reynaert and F.~Verboven.
\newblock Improving the performance of random coefficients demand models: the
  role of optimal instruments.
\newblock \emph{Journal of Econometrics}, 179\penalty0 (1):\penalty0 83--98,
  2014.

\bibitem[Skrainka and Judd(2011)]{skrainka2011high}
B.~S. Skrainka and K.~L. Judd.
\newblock High performance quadrature rules: How numerical integration affects
  a popular model of product differentiation.
\newblock \emph{Available at SSRN 1870703}, 2011.

\bibitem[Varadhan and Roland(2008)]{varadhan2008simple}
R.~Varadhan and C.~Roland.
\newblock Simple and globally convergent methods for accelerating the
  convergence of any em algorithm.
\newblock \emph{Scandinavian Journal of Statistics}, 35\penalty0 (2):\penalty0
  335--353, 2008.

\end{thebibliography}
\pagebreak

\begin{table}[h!]
    \centering
    \caption{Notation.}
    \begin{tabular}{c l}
    \toprule
    Variable & Definition \\
    \midrule
    $t$ & Markets in range   $\{1, 2, 3, ..., T\}$\\\addlinespace[5pt]
    $F_{t}$ & The number of firms in market $t$. \\\addlinespace[5pt]
    $f$ & Firm $f$ in range $\{1, ..., F_{t}\}$\\\addlinespace[5pt]
    $j$ & Product $j$ in range $\{1, 2, 3, ..., N_{t}$ \\\addlinespace[5pt]
    $\mathcal{J}_{t}$ & Set of products in market $t$  \\\addlinespace[5pt]
    $\mathcal{J}_{ft}$ & Set of products in firm $f$ and market $t$  \\\addlinespace[5pt]
    $N_{t}$ & Number of products in market $t$  \\\addlinespace[5pt]
    $N_{ft}$ & Number of products in firm $f$ and market $t$  \\\addlinespace[5pt]
    $x_j$ & Observed characteristics of product $j$  \\\addlinespace[5pt]
    $\beta_i$ & Consumer $i$'s taste parameter  \\\addlinespace[5pt]
    $p_j$ & Price of product $j$ \\\addlinespace[5pt]
    $c_j$ & Marginal cost of product $j$ \\\addlinespace[5pt]
    $\xi_j$ & Unobserved characteristics of product $j$\\\addlinespace[5pt]
    $\epsilon_{ij}$ & Distribution of consumer preferences around the mean\\\addlinespace[5pt]
    $\theta$ & A vector of all parameters.\\
    \bottomrule
    \end{tabular}
    \label{model:var_def}
    \end{table}
    
    \begin{table}\centering
    \caption{RMSE of simulations.}
      \begin{tabular}{lcccccc}
\toprule
&\multicolumn{3}{c}{GMM}&\multicolumn{3}{c}{MLE}\\
\cmidrule(lr){2-4}
\cmidrule(lr){5-7}
& $\alpha$ & $\sigma_{x}$ & $\sigma_{price}$ & $\alpha$ & $\sigma_{x}$ & $\sigma_{price}$\\
\midrule  
\textbf{No covariance} & 0.42 & 0.34 & 0.13 & 0.20 & 0.23 & 0.10\\\addlinespace[5pt]
\textbf{Low covariance} & 0.41 & 0.27 & 0.13 & 0.28 & 0.20 & 0.11\\\addlinespace[5pt]
\textbf{High covariance} & 0.40 & 0.29 & 0.14 & 0.43 & 0.20 & 0.11\\\addlinespace[5pt]
\textbf{Laplace, no covariance} & 0.46 & 0.34 & 0.13 & 0.22 & 0.23 & 0.11\\\addlinespace[5pt]
\textbf{Laplace, low covariance} & 0.38 & 0.27 & 0.13 & 0.33 & 0.21 & 0.12\\\addlinespace[5pt]
\textbf{Supply misspecification} & 0.42 & 0.32 & 0.14 & 0.99 & 0.41 & 0.13\\\addlinespace[5pt]
\textbf{Ownership misspecification} & 0.39 & 0.32 & 0.12 & 0.34 & 0.21 & 0.11\\
\bottomrule
\end{tabular}

    \vspace{5pt}
    \caption*{\textit{Notes} This table presents the root mean squared error of different simulation scenarios. Each column is a combination of an estimator and a parameter. We use 1000 simulations to generate each row in this table. Note that the supply misspecification and the ownership misspecification scenarios also have low covariance.}
        \label{tab:rmse_summary}
      
    \end{table}
    
    \begin{table}\centering
       \caption{Mean bias of simulations.}
      \begin{tabular}{lcccccc}
\toprule
&\multicolumn{3}{c}{GMM}&\multicolumn{3}{c}{MLE}\\
\cmidrule(lr){2-4}
\cmidrule(lr){5-7}
& $\alpha$ & $\sigma_{x}$ & $\sigma_{price}$ & $\alpha$ & $\sigma_{x}$ & $\sigma_{price}$\\
\midrule  
\textbf{No covariance} & 0.12 & -0.09 & 0.03 & 0.03 & 0.01 & -0.02\\\addlinespace[5pt]
\textbf{Low covariance} & 0.10 & -0.04 & 0.01 & 0.03 & 0.01 & -0.02\\\addlinespace[5pt]
\textbf{High covariance} & 0.04 & -0.08 & 0.01 & 0.05 & -0.00 & -0.03\\\addlinespace[5pt]
\textbf{Laplace, no covariance} & 0.11 & -0.08 & 0.02 & 0.03 & -0.00 & -0.01\\\addlinespace[5pt]
\textbf{Laplace, low covariance} & 0.08 & -0.06 & 0.02 & 0.05 & 0.00 & -0.02\\\addlinespace[5pt]
\textbf{Supply misspecification} & 0.20 & 0.08 & -0.01 & 0.94 & 0.30 & -0.05\\\addlinespace[5pt]
\textbf{Ownership misspecification} & -0.05 & -0.20 & 0.03 & -0.00 & -0.09 & -0.02\\
\bottomrule
\end{tabular}

      \vspace{5pt}
       \caption*{\textit{Notes} This table presents the root mean bias of different simulation scenarios. Each column is a combination of an estimator and a parameter. We use 1000 simulations to generate each row in this table. Note that the supply misspecification and the ownership misspecification scenarios also have low covariance.}
       \label{tab:mean_bias_summary}
    \end{table}
    
    \begin{table}\centering
        \caption{Mean standard error values.}
      \begin{tabular}{lcccccc}
\toprule
&\multicolumn{3}{c}{GMM}&\multicolumn{3}{c}{MLE}\\
\cmidrule(lr){2-4}
\cmidrule(lr){5-7}
& $\alpha$ & $\sigma_{x}$ & $\sigma_{price}$ & $\alpha$ & $\sigma_{x}$ & $\sigma_{price}$\\
\midrule  
\textbf{No covariance} & 0.30 & 0.30 & 0.13 & 0.20 & 0.22 & 0.12\\\addlinespace[5pt]
\textbf{Low covariance} & 0.28 & 0.27 & 0.22 & 0.23 & 0.19 & 0.12\\\addlinespace[5pt]
\textbf{High covariance} & 0.28 & 0.29 & 0.16 & 0.30 & 0.20 & 0.14\\\addlinespace[5pt]
\textbf{Laplace, no covariance} & 0.30 & 0.32 & 0.14 & 0.20 & 0.22 & 0.13\\\addlinespace[5pt]
\textbf{Laplace, low covariance} & 0.28 & 0.28 & 0.15 & 0.25 & 0.20 & 0.13\\\addlinespace[5pt]
\textbf{Supply misspecification} & 0.29 & 0.30 & 0.16 & 0.20 & 0.21 & 0.10\\\addlinespace[5pt]
\textbf{Ownership misspecification} & 0.26 & 0.24 & 0.10 & 0.30 & 0.19 & 0.13\\
\bottomrule
\end{tabular}

        \vspace{5pt}
        \caption*{\textit{Notes} This table presents the standard errors of different simulation scenarios. Each column is a combination of an estimator and a parameter. We use 1000 simulations to generate each row in this table. Note that the supply misspecification and the ownership misspecification scenarios also have low covariance. In the GMM sigma price column, we omit approximately $10\%$ of the simulations because the standard error calculation for these simulations were producing numerically infeasible values.} 
        \label{tab:mean_se}
    \end{table}
    \begin{table}\centering
    \caption{Fraction of 95\% CI that include the true values.}
      \begin{tabular}{lcccccc}
\toprule
&\multicolumn{3}{c}{GMM}&\multicolumn{3}{c}{MLE}\\
\cmidrule(lr){2-4}
\cmidrule(lr){5-7}
& $\alpha$ & $\sigma_{x}$ & $\sigma_{price}$ & $\alpha$ & $\sigma_{x}$ & $\sigma_{price}$\\
\midrule    
\textbf{No covariance} & 0.86 & 0.92 & 0.79 & 0.94 & 0.93 & 0.94\\\addlinespace[5pt]
\textbf{Low covariance} & 0.89 & 0.95 & 0.77 & 0.93 & 0.94 & 0.94\\\addlinespace[5pt]
\textbf{High covariance} & 0.84 & 0.93 & 0.72 & 0.88 & 0.94 & 0.95\\\addlinespace[5pt]
\textbf{Laplace, no covariance} & 0.86 & 0.93 & 0.79 & 0.94 & 0.95 & 0.92\\\addlinespace[5pt]
\textbf{Laplace, low covariance} & 0.88 & 0.95 & 0.77 & 0.91 & 0.94 & 0.90\\\addlinespace[5pt]
\textbf{Supply misspecification} & 0.93 & 0.97 & 0.70 & 0.01 & 0.64 & 0.86\\\addlinespace[5pt]
\textbf{Ownership misspecification} & 0.83 & 0.81 & 0.80 & 0.89 & 0.91 & 0.93\\
\bottomrule
\end{tabular}

    \vspace{5pt}
    \caption*{\textit{Notes} This table presents the coverage of different estimators. Each column is a combination of an estimator and a parameter, and the entries are the fraction of the simulations in which the truth lay within the 95\% confidence intervals produced by each estimator. We use 1000 simulations to generate each row in this table. Note that the supply misspecification and the ownership misspecification scenarios also have low covariance. 
    }
        \label{tab:percent_in_ci}
    
    \end{table}

    
    
    \begin{table}\centering\footnotesize
        \caption{BLP automobile data parameter estimates with standard errors.}
      
\begin{tabular}{llrrrr}
    \toprule
     & Variable & \textbf{\shortstack{BLP}} & \textbf{\shortstack{AGS}} & \textbf{\shortstack{CG}} & \textbf{\shortstack{MLE}}\\
     \midrule
     \addlinespace[5pt]
     Term on Price ($\alpha$) & $\ln(y-p)$ & 43.501 & 42.870 & 45.898 & 45.227\\
     & & (6.43) & (5.56) & (11.75) & (0.77)\\
     \addlinespace[10pt]
     Standard deviations & Constant & 3.612 & 2.522 & 2.962 & 1.924\\
     & & (1.49) & (2.54) & (1.64) & (0.16)\\
     \addlinespace[5pt]
     & HP/Weight & 4.628 & 3.525 & 1.388 & 4.611\\
     & & (1.89) & (2.84) & (2.11) & (0.28)\\
     \addlinespace[5pt]
     & Air & 1.818 & 4.167 & 1.424 & 1.991\\
     & & (1.70) & (1.41) & (0.43) & (0.23)\\
     \addlinespace[5pt]
     & MP\$ & 1.050 & 0.393 & 0.072 & 0.299\\
     & & (0.27) & (0.28) & (1.00) & (0.04)\\
     \addlinespace[5pt]
     & Size & 2.056 & 1.937 & 0.231 & 2.168\\
     & & (0.58) & (0.60) & (3.84) & (0.11)\\
     \addlinespace[10pt]
     Means & Constant & -7.061 & -7.728 & -6.679 & -5.439\\
     & & (0.94) & (1.16) & (1.30) & (0.33)\\
     \addlinespace[5pt]
     & HP/Weight & 2.883 & 4.620 & 2.774 & 3.175\\
     & & (2.02) & (1.13) & (0.83) & (0.35)\\
     \addlinespace[5pt]
     & Air & 1.521 & -1.227 & 0.572 & 0.385\\
     & & (0.89) & (1.38) & (0.35) & (0.16)\\
     \addlinespace[5pt]
     & MP\$ & -0.122 & 0.293 & 0.340 & 0.086\\
     & & (0.32) & (0.16) & (0.10) & (0.06)\\
     \addlinespace[5pt]
     & Size & 3.460 & 3.992 & 3.920 & 3.327\\
     & & (0.61) & (0.35) & (0.32) & (0.17)\\
     \addlinespace[10pt]
     Cost Side Parameters & Constant & 0.952 & 2.751 & 2.780 & 2.327\\
     & & (0.19) & (0.08) & (0.10) & (0.04)\\
     \addlinespace[5pt]
     & $\ln(\text{HP/Weight})$ & 0.477 & 0.812 & 0.731 & 0.495\\
     & & (0.06) & (0.06) & (0.07) & (0.03)\\
     \addlinespace[5pt]
     & Air & 0.619 & 0.430 & 0.528 & 0.601\\
     & & (0.04) & (0.05) & (0.04) & (0.02)\\
     \addlinespace[5pt]
     & $\ln(y-p)$ & -0.415 & -0.610 & -0.651 & -0.377\\
     & & (0.06) & (0.05) & (0.07) & (0.04)\\
     \addlinespace[5pt]
     & $\ln(\text{Size})$ & -0.046 & -0.352 & -0.472 & 0.028\\
     & & (0.08) & (0.11) & (0.12) & (0.06)\\
     \addlinespace[5pt]
     & Trend & 0.019 & 0.027 & 0.018 & 0.015\\
     & & (0.00) & (0.00) & (0.00) & (0.00)\\
     \bottomrule
\end{tabular}
        \vspace{5pt}
        \caption*{\textit{Notes} This table presents parameter estimates of different estimators on the BLP automobile data. Column (1) has the original estimates, column (2) is from the AGS data, (3) is the best practice estimates from the CG paper, (4) is the parameter estimates from the MLE estimator. Standard errors are shown parentheses.}
        \label{tab:blp95_param_se_estimates}
    \end{table}
\clearpage
\appendix
\section{Appendix}
\subsection{Deriving and simplifying the log-likelihood}\label{sec:mle_algebra}
In this section of the appendix, we are concerned with deriving terms and simplifying the unconcentrated log likelihood. We first show how the parameters $(\Sigma,\beta,\gamma)$ may be concentrated out, and then provide an analysis of the Jacobian term in the concentrated log likelihood.  

\subsubsection{Concentrating out $\Sigma$}\label{sec:part_2_simp}
Define the following equalities for a given dataset ranging with observations (product-markets) ranging from 1 to $m$:

\begin{equation}
\begin{aligned}
    \sum_{i=1}^{m}\xi_i^2 = \sum_{i=1}^{m}(\delta_i - x_i \beta)^2 = (\delta - X\beta)^T(\delta - X\beta)  = \xi^T \xi\\
    \sum_{i=1}^{m}u_i^2 =\sum_{i=1}^{m}(c_i - w_i \gamma)^2  = (C - W\gamma)^T(C - W\gamma)  = u^T u\\
  \end{aligned}
\end{equation}
Recall Part II of the log-likelihood is given by: 
\begin{equation}
  \begin{aligned}
  =& ([\delta_j \;  c_j] - E[\delta_j \;  c_j | x_{t}, w_{t}])\Sigma^{-1} (\begin{bmatrix}\delta_j \\  c_j\end{bmatrix}- E[\delta_j \;  c_j | x_{t}, w_{t}]^T)\\
  =&([\delta_j \;  c_j]- E[\delta_j \;  c_j | x_{t}, w_{t}])\begin{bmatrix}\sigma_{\xi}^2 \; \sigma_{\xi, u} \\  \sigma_{\xi, u} \; \sigma_u^2 \\ \end{bmatrix}^{-1} (\begin{bmatrix}\delta_j \\ c_j\end{bmatrix}- E[\delta_j \;  c_j | x_{t}, w_{t}]^T) \\
  \end{aligned}
\end{equation}

\noindent where for simplicity we index observations by $j$ instead of $jt$. The maximum likelihood estimate of $\Sigma$, holding the other parameters fixed, is the sample variance-covariance matrix of the error terms. 

\begin{equation}
  \begin{aligned}
    = & \begin{bmatrix}\xi_j \; u_j\end{bmatrix}\begin{bmatrix}
    \sum_{i=1}^{m}\xi_i^2 & \sum_{i=1}^{m}\xi_iu_i \\
    \sum_{i=1}^{m}\xi_iu_i & \sum_{i=1}^{m}u_i^2
  \end{bmatrix}^{-1}
     \begin{bmatrix}\xi_j \\  u_j\end{bmatrix}\\
   =&\frac{1}{\sum_{i=1}^{m}\xi_i^2 \sum_{i=1}^{m}u_i^2 -(\sum_{i=1}^{m}\xi_i u_i)^2}\begin{bmatrix}\xi_j \; u_j\end{bmatrix}\begin{bmatrix}
   \sum_{i=1}^{m}u_i^2 & -\sum_{i=1}^{m}\xi_i u_i \\
   -\sum_{i=1}^{m}\xi_iu_i & \sum_{i=1}^{m}\xi_i^2
 \end{bmatrix}
    \begin{bmatrix}\xi_j \\ u_j\end{bmatrix}\\
      =&\frac{\xi_j^2\sum_{i=1}^{m}u_i^2   -2u_j \xi_j\sum_{i=1}^{m}\xi_i u_i   + u_j^2\sum_{i=1}^{m}\xi_i^2 }{\sum_{i=1}^{m}\xi_i^2 \sum_{i=1}^{m}u_i^2 -(\sum_{i=1}^{m}\xi_i u_i)^2}
      \\
    \end{aligned}
\end{equation}

\noindent Next, we can aggregate the likelihood across all data points:

\begin{equation}
\begin{aligned}
  =& \sum_{j=1}^{m}\left( \frac{\xi_j^2\sum_{i=1}^{m}u_i^2   -2u_j \xi_j\sum_{i=1}^{m}\xi_i u_i   + u_j^2\sum_{i=1}^{m}\xi_i^2 }{\sum_{i=1}^{m}\xi_i^2 \sum_{i=1}^{m}u_i^2 -(\sum_{i=1}^{m}\xi_i u_i)^2}\right)
  \\
  =& \frac{\sum_{j=1}^{m}\xi_j^2\sum_{i=1}^{m}u_i^2   -2\sum_{j=1}^{m}u_j \xi_j\sum_{i=1}^{m}\xi_i u_i   + \sum_{j=1}^{m}u_j^2\sum_{i=1}^{m}\xi_i^2 }{\sum_{i=1}^{m}\xi_i^2 \sum_{i=1}^{m}u_i^2 -(\sum_{i=1}^{m}\xi_i u_i)^2}
  \\
  =& \frac{2\sum_{j=1}^{m}\xi_j^2\sum_{i=1}^{m}u_i^2   -2(\sum_{j=1}^{m}u_j \xi_j)^2}{\sum_{i=1}^{m}\xi_i^2 \sum_{i=1}^{m}u_i^2 -(\sum_{i=1}^{m}\xi_i u_i)^2}
  \\
  =& 2
  \\
\end{aligned}
\end{equation}
\subsubsection{Estimating the linear parameters}\label{sec:variance_estimator}

This section is concerned with the estimation of Part I of eq. \ref{eq:likelihood}. As mentioned before, $u$ and $\xi$ have a bivariate normal distribution with arbitrary correlation. 
\noindent The local objective function that we are concerned with for this part is is:
\begin{equation}
  \begin{aligned}
    \min_{\gamma, \beta}\log(|\hat{\Sigma}|)&=&\min_{\gamma, \beta}\log \bigg(\bigg|\begin{bmatrix}
      \sum_{i=1}^{m}\xi_i^2 & \sum_{i=1}^{m}\xi_iu_i \\
      \sum_{i=1}^{m}\xi_iu_i & \sum_{i=1}^{m}u_i^2 \end{bmatrix}\bigg|\bigg)\\
      &=&\min_{\gamma, \beta}\log\bigg(\sum_{i=1}^{m}\xi_i^2\sum_{i=1}^{m}u_i^2 - \left( \sum_{i=1}^{m}\xi_iu_i\right)^2\bigg)\\
  \end{aligned}
\end{equation}

\noindent where $|\cdot|$ is the determinant. Since $\log(\cdot)$ is monotonic, the objective above is equivalent to:

\begin{equation}
  \begin{aligned}
      \min_{\gamma, \beta}\log(|\Sigma|)&=&\min_{\gamma, \beta}\left(\sum_{i=1}^{m}\xi_i^2\sum_{i=1}^{m}u_i^2 -  \left(\sum_{i=1}^{m}\xi_iu_i\right)^2\right)\\
  \end{aligned}
\end{equation}

\noindent Next, we can write out the objective in matrix form and find the FOCs:

\small
\begin{equation}
  \begin{aligned}
    O = (\delta - X\beta)^T(\delta - X\beta)(C - W\gamma)^T(C - W\gamma) - [(\delta - X\beta)^T(C - W\gamma)]^2\\
  \end{aligned}
\end{equation}
\normalsize
Taking a derivative in $\beta$:
\begin{equation}\label{eq:als_1}
  \begin{aligned}
    \frac{\partial O}{\partial \beta} = -2X^T(\delta - X\beta)(C - W\gamma)^T(C - W\gamma) - 2 X^T (C - W\gamma)(C - W\gamma)^T(\delta - X\beta)\\
    0 =  X^T(\delta - X\beta)u^T u - 2 X^T u u^T (\delta - X\beta)\\
    X^T X \beta u^T u + X^T u u^T X \beta = X^T \delta u^T u + X^T u u^T \delta\\
    (X^T Xu^T u + X^T u u^T X)\beta = X^T\delta u^T u + X^T u u^T \delta\\
    (X^T Xu^T u + X^T u u^T X)\beta = X^T\delta u^T u + X^T u u^T \delta\\
    \hat{\beta} = (X^T Xu^T u + X^T u u^T X)^{-1}(X^T\delta u^T u + X^T u u^T \delta)\\
  \end{aligned}
\end{equation}

Similarly, for $\gamma$:
\begin{equation}\label{eq:als_2}
  \begin{aligned}
    \frac{\partial O}{\partial \gamma} = -2W^T(C - W\gamma)(\delta - X\beta)^T(\delta - X\beta) - 2 W^T (\delta - X\beta)(\delta - X\beta)^T(C - W\gamma)\\
    \hat{\gamma} = (W^T W\xi^T \xi + W^T \xi \xi^T W)^{-1}(W^TC \xi^T \xi + W^T \xi \xi^T C)\\
  \end{aligned}
\end{equation}

Equations \ref{eq:als_1} and \ref{eq:als_2} are dependent on each other. To solve for $\hat{\beta}$ and $\hat{\gamma}$, we use Alternating Least Squares (ALS). 

\subsubsection{The Jacobian in the Likelihood Function} \label{sec:jacobian}
To derive the Jacobian component in the likelihood function, we will use a triangular system to estimate the total derivatives: (1) Use the supply side first order conditions (FOCs) to get derivatives with respect to price, mean utility, and costs, (2) get the price total derivatives using the Implicit Function Theorem (IFT) on the FOCs, (3) get the share total derivatives using the price total derivatives and the share partial derivatives.

\paragraph{Getting the supply side FOCs.} Using the Bertrand-Nash equilibrium, we assume that firms act using the following objective:
\begin{equation}
 \max_{p_j} \pi^{f,t} = \sum_{j \in N_{ft}} (p_{j,t} - c_{j,t})s_{j,t}.
\end{equation}
For the rest of this derivation, we  drop the market notation (all math below is for a single market). The product specific first order condition is:

\begin{equation}
  \begin{aligned}
     F_j^f = \frac{\partial \pi^f}{\partial p_j} = s_j + \sum_{k \in F(j)} (p_k - c_k) \frac{\partial s_k}{\partial p_j} \\
  \end{aligned}
\end{equation}

Next, we take the derivative of the FOCs with respect to price, mean utility, and costs:
\begin{align}
 \frac{\partial F_j^f}{\partial p_l} &=   \frac{\partial s_j}{\partial p_l} + \underbrace{\frac{\partial s_k}{\partial p_l}}_\text{if $k \in F(j)$} + \underbrace{\sum_{k \in F(j)} (p_k - c_k) \frac{\partial s_k}{\partial p_j \partial p_l}}_\text{$\Xi^{p, p}_{j, l}$}   \\
  \frac{\partial F_j^f}{\partial \delta_l} &= \frac{\partial s_j}{\partial \delta_l} + \underbrace{\sum_{k \in F(j)} (p_k - c_k) \frac{\partial s_k}{\partial p_j \partial \delta_l}}_\text{$\Xi^{p, \delta}_{j, l}$}  \\
 \frac{\partial F_j^f}{\partial c_l} &= \underbrace{-\frac{\partial s_l}{\partial p_j}}_\text{if $l \in F(j)$}.
\end{align}
We can write this in matrix notation, where the partial derivative of vector $y$ with respect to vector $x$ can be denoted as $J^{y, x}$. Additionally, as noted below the terms above, the $\Xi$ terms are sparse Hessians (sparsity comes from the $k\in F(j)$). With these two notational changes in mind, we can write the above equations in a simpler form:\footnote{Note that we can do this since $J^{s, p}$ is symmetric.}
\begin{align}
     J^{F, p} &=   J^{s, p} + O \circ J^{s, p} + \Xi^{p,p}\\
     J^{F, \delta} &= J^{s, \delta} + \Xi^{p, \delta}\\
     J^{F, c} &= -O \circ J^{s, p},
\end{align}
where $\circ$ is the Hadamard (i.e., element-wise) matrix-product operator. This gives us the variables of interest. Formulae for the Hessian tensors in $\Xi$ will be derived later in the last paragraph of this section.

\paragraph{Getting the price derivatives.} To get the price derivatives we will use the IFT on the first derivative of the supply side equation as follows:
\begin{equation}
     \frac{\partial \pi_{ft}}{\partial p_{jt}} = F_{fjt}(\delta_{t},p_{t},c_{t}) \equiv \sum_{j' \in N_{ft}}(p_{j't} - c_{j't})\frac{\partial s_{j't}}{\partial p_{jt}} + s_{jt} 1_{j \in N_{ft}} = 0.
\end{equation}
The IFT implies:
\begin{equation}
  \begin{aligned}
        \begin{bmatrix}
          \frac{d p_t}{d \delta_t} \; | \;
          \frac{d p_t}{d c_t}
        \end{bmatrix} = -(J^{F, p}_t)^{-1} \begin{bmatrix}
        J^{F, \delta}_t \; | \;
        J^{F, c}_t
      \end{bmatrix}
  \end{aligned}
\end{equation}
Applying the IFT to the $\mathcal{J}_t \times 2\mathcal{J}_t$ matrix formed by stacking the
matrices next to each other gives:
\begin{equation}\small
  \begin{aligned}
    \begin{bmatrix}
      \frac{d p_t}{d \delta_t} \; | \;
      \frac{d p_t}{d c_t}
    \end{bmatrix} & =
    -\left(\frac{\partial F_t}{\partial p_t}\right)^{-1}
    \begin{bmatrix}
            \frac{\partial F_t}{\partial \delta_t} \;
            \frac{\partial F_t}{\partial c_t}
    \end{bmatrix} \\
    & = - \begin{bmatrix}
    J^{s, p}_t + O_t \circ J^{s, p}_t + \Xi^{p,p}_t
        \end{bmatrix}^{-1}
        \begin{bmatrix}
            J^{s, \delta}_t + \Xi^{p, \delta}_t \; | \;
            -O_t \circ J^{s, p}_t
        \end{bmatrix}
  \end{aligned},
\end{equation}
which is a $\mathcal{J}_t \times (2\mathcal{J}_t)$ matrix with of derivatives of all the prices (rows) in $\delta$ (first set of $\mathcal{J}_t$ columns) and $c$ (last set of $\mathcal{J}_t$ columns) respectively.

\paragraph{Share derivatives.} Using the derived values above, we can calculate the last two entries of the Jacobian matrix of interest, the full derivatives of shares with respect to mean utility and costs:
\begin{align}
    \frac{d s_t}{d \delta_t} &= \frac{\partial s_t}{\partial \delta_t} + \frac{\partial s_t}{\partial p_t}\frac{d p_t}{d \delta_t}\\
    \frac{d s_t}{d c_t} &= \frac{\partial s_t}{\partial p_t}\frac{d p_t}{d c_t}\\
\end{align}
where the derivatives of $p$ w.r.t. to exogenous variables are given by the IFT above. Now, we have a way of getting the Jacobian matrix for the likelihood:

\begin{equation}
\begin{aligned}
  J = \begin{bmatrix} \frac{d s}{d \delta} \;  \frac{d s}{d c}\\
  \frac{d p}{d \delta} \;  \frac{d p}{d c}\\
\end{bmatrix}
\end{aligned}
\end{equation}
\paragraph{Deriving the Hessians.}\label{sec:hessian}
From above, we need to calculate the following Hessian terms: $H^{s, p, p}_{k, j, l} = \frac{\partial s_k}{\partial s_j \partial s_l}$ and $H^{s, p, \delta}_{k, j, l} = \frac{\partial s_k}{\partial s_j \partial \delta_l}$ for each market. 
We can write shares for product $k$ as:
\begin{align}
  s_k & = \int \frac{e^{V_{ki}}}{1 + \sum_{w \in \mathcal{J}_t} e^{V_{wi}}} d\epsilon_{ki}\\
  &= \int s_{ki}d \epsilon_{ki},
\end{align}
which we can use to get the derivative of shares with respect to price:
\begin{equation}
  \frac{\partial s_k}{\partial p_j} = \int \frac{\partial V_{ji}}{\partial p_j} (\underbrace{s_{ki}}_\text{if $k=j$} - s_{ki}*s_{ji}) \; d\epsilon_{ki}.
\end{equation}
Taking a second derivative with respect to prices:
\begin{equation}
  \frac{\partial s_k}{\partial p_j \partial p_l} = \int \frac{\partial V_{ji}}{\partial p_j} \frac{\partial V_{li}}{\partial p_l} (\underbrace{s_{ki}}_\text{if $k=j=l$} - \; \; \underbrace{s_{ki}s_{li}}_\text{if $k=j$} \; \; \underbrace{-s_{ki}s_{ji}}_\text{if $j=l$} \; \; \underbrace{-s_{ki}s_{ji}}_\text{if $k=l$}  + 2 s_{ki}s_{ji}s_{li}) \; d\epsilon_{ki}.
\end{equation}
$H^{s, p, \delta}_{k, j, l}$ can be derived similarly:
\begin{equation}
  \frac{\partial s_k}{\partial p_j \partial \delta_l} = \int \frac{\partial V_{ji}}{\partial p_j} \frac{\partial V_{li}}{\partial \delta_l} (\underbrace{s_{ki}}_\text{if $k=j=l$} - \; \; \underbrace{s_{ki}s_{li}}_\text{if $k=j$} \; \; \underbrace{-s_{ki}s_{ji}}_\text{if $j=l$} \; \; \underbrace{-s_{ki}s_{ji}}_\text{if $k=l$}  + 2 s_{ki}s_{ji}s_{li}) \; d\epsilon_{ki}.
\end{equation}

\subsection{Integrating market shares}\label{sec:quadrature}
Accurate integration is crucial for optimization. One option is Monte Carlo integration, which uses randomly drawn nodes with equal weights of $1/draws$. 
A more precise and computationally efficient way is
Gauss-Hermite quadrature, which deals with integrals of the form $\int e^{-x^2}f(x) \approx \sum_{i=1}^n w_i f(x_i)$. 
For the multinomial logit setup of BLP, \cite{skrainka2011high} propose the quadrature setup:
\begin{equation}
  \begin{aligned}
    s_{jt} = \int \frac{e^{\delta_{jt} + x_{jt} \sigma_{\beta} v_{it}}}{1 + \sum_{j' \in \mathcal{J}_t} e^{\delta_{j't} + x_{j't}  \sigma_{\beta} v_{it}}} f(v_{it}) dv_{it}\\
  \end{aligned}
\end{equation}
where $v_{it}$ is a normal variable with an identity for diagonal. The
Gauss-Hermite estimation of this integral can calculated once $f(v_{it})$ is
replaced with the normal distribution formula and a change of variables setup is
used to satisfy the form of $\int e^{-x^2}f(x)$:
\begin{equation}
    s_{jt} \approx \pi^{-\frac{K}{2}} \sum_{i=1}^n w_i \frac{e^{\delta_{jt} + x_{jt} \sigma_{\beta}v_{it}}}{1 + \sum_{j' \in \mathcal{J}_t} e^{\delta_{j't} + x_{j't}  \sigma_{\beta} v_{it}}}.
\end{equation}

\subsection{Standard Error Estimation} \label{sec:std_err}

We refer the reader to~\cite{conlon2020best} for GMM standard error estimation. For MLE, we have:
\begin{equation}
    \sqrt{n}(\hat{\Theta} - \Theta_0) \xrightarrow[]{} N(0, I^{-1})
\end{equation}
for $I$ the Fisher Information matrix:
\begin{equation}
  \begin{aligned}
    I = -E\left[  \frac{\partial^2 \ell(\mathcal{D} ; \Theta)}{\partial \Theta \partial \Theta'}\right]\
  \end{aligned}
\end{equation} 
We estimate this as follows:
\begin{equation}
  \begin{aligned}
    \widehat{I} = -\frac{\partial^2}{\partial \Theta \partial \Theta'}\left(
    \frac{1}{T} \sum_t  \ell(\mathcal{D}_t ; \Theta)\right)
  \end{aligned}
\end{equation} 
where we have interchanged expectation and derivative, and replaced the expectation with its empirical analog.
We compute derivatives using automatic differentiation.

\end{document}